\documentclass[aps,reprint,superscriptaddress,
nofootinbib,showpacs,amsmath,amssymb]{revtex4-1}

\usepackage{latexsym,graphicx,slashed,bm,dcolumn}
\usepackage{hyperref}

\newcommand{\be}[1]{\begin{equation}\label{#1}}
\newcommand{\beq}{\begin{equation}}
\newcommand{\eeq}{\end{equation}}
\def\ee{\end{equation}}
\newcommand{\beqn}[1]{\begin{eqnarray}\label{#1}}
\newcommand{\eeqn}{\end{eqnarray}}

\newcommand{\dub}[2]{\left(\begin{array}{c}{#1}\\{#2}
\end{array}\right)}

\newcommand{\mat}[4]{\left(\begin{array}{cc}{#1}&{#2}\\{#3}&{#4}
\end{array}\right)}
\newcommand{\matr}[9]{\left(\begin{array}{ccc}{#1}&{#2}&{#3}\\
{#4}&{#5}&{#6}\\{#7}&{#8}&{#9}\end{array}\right)}

\renewcommand{\to}{\rightarrow}

\def\ov{\overline}

\def\lsim{\raise0.3ex\hbox{$\;<$\kern-0.75em\raise-1.1ex
\hbox{$\sim\;$}}}
\def\gsim{\raise0.3ex\hbox{$\;>$\kern-0.75em\raise-1.1ex
\hbox{$\sim\;$}}}
\def\cal{\mathcal}
\def\cF{{\cal F}}

\def\cM{{\cal M}}

\begin{document}

\title{The CKM unitarity problem:  A trace of  new physics at the TeV scale?   }

\author{Benedetta~Belfatto}
\affiliation{Gran Sasso Science Institute,  67100, L'Aquila, Italy} 
\affiliation{INFN, Laboratori Nazionali del Gran Sasso, 67010 Assergi,  L'Aquila, Italy}
\author{Revaz~Beradze}
\affiliation{Ivane Javakhishvili Tbilisi State University, Chavchavadze Av.~3, 0179 Tbilisi, Georgia } 
\affiliation{Dipartimento di Fisica e Chimica, Universit\`a di L'Aquila, 67100 Coppito, L'Aquila, Italy} 
\author{Zurab~Berezhiani}
\email{E-mail: zurab.berezhiani@lngs.infn.it}
\affiliation{Dipartimento di Fisica e Chimica, Universit\`a di L'Aquila, 67100 Coppito, L'Aquila, Italy} 
\affiliation{INFN, Laboratori Nazionali del Gran Sasso, 67010 Assergi,  L'Aquila, Italy}


\begin{abstract} 
After the recent high precision determinations of $V_{us}$ and $V_{ud}$, the first row of the CKM matrix 
shows more than $4\sigma$ deviation from unitarity. 
Two possible scenarios beyond the Standard Model can be investigated in order to fill the gap. 
If a 4th quark $b'$ participates in the mixing, with  $\vert V_{ub'} \vert \sim0.04$, 
then its mass should be no more than 6 TeV or so.  
A different solution can come from the introduction of the gauge horizontal family symmetry 
acting between the lepton families 
and spontaneously broken at the scale of about 6 TeV. 
Since the gauge bosons of this symmetry contribute to muon decay in interference with Standard Model,
 the Fermi constant is slightly smaller than the muon decay constant so that unitarity is recovered. 
Also the neutron lifetime problem, that is about $4\sigma$ discrepancy between 
 the neutron lifetimes measured  in beam and trap experiments, is discussed in the light of the  
 these determinations of the CKM matrix elements.
\end{abstract}


\maketitle




\noindent {\bf 1.} 
The Standard Model (SM)  contains three fermion families in the identical representations 
of the gauge symmetry $SU(3)\times SU(2)\times U(1)$ of strong and electroweak interactions.  
One of its fundamental predictions 
is the unitarity of the Cabibbo-Kobayashi-Maskawa (CKM) matrix 
of quark mixing in charged current
\be{CKM} 
V_{\rm CKM} = \matr{V_{ud} }{V_{us} } {V_{ub} }  {V_{cd} } {V_{cs} } {V_{cb} }  {V_{dd} }{V_{ts} }{V_{tb} }   .
\ee
Deviation from the CKM unitarity can be a signal of new physics beyond the Standard Model (BSM).  
The experimental precision and control of theoretical uncertainties in the 
determination of the elements in the first row of $V_{\rm CKM}$ 
are becoming sufficient for testing the condition
\be{unitarity} 
\vert V_{ud} \vert^2 + \vert V_{us} \vert^2 + \vert V_{ub} \vert^2 = 1 \, . 
\ee 
Since $\vert V_{ub} \vert \simeq 0.004$ is very small, its  contribution 
is negligible and  (\ref{unitarity})  reduces essentially to the check of the Cabibbo mixing:   
$\vert V_{us} \vert = \sin\theta_C$, $\vert V_{ud} \vert = \cos\theta_C$ 
and $\vert V_{us}/V_{ud} \vert = \tan\theta_C$.
In essence, this is the universality test for the $W$-boson coupling  
$(g/\sqrt2) W^+_\mu J^\mu_L + {\rm h.c.}$  to the relevant part of the charged 
left-handed current 
\be{explicit} 
J_L^\mu = V_{ud} \ov{u_L} \gamma^\mu d_L + V_{us} \ov{u_L} \gamma^\mu s_L 
+  \ov{\nu_{e} } \gamma^\mu e_L + \ov{\nu_{\mu}} \gamma^\mu \mu_L 
\ee
For energies smaller than $W$-boson mass 
 this coupling gives rise to the effective current $\times$ current interactions
 \be{ud}
- \frac{4G_F}{\sqrt2} \, \ov{u_L}  \big( V_{ud} \gamma_\mu  d_L + V_{us} \gamma_\mu  s_L\big) 
\big(\ov{e_L} \gamma^\mu \nu_e  + \ov{\mu_L} \gamma^\mu \nu_\mu \big) 
\ee
which are responsible for leptonic decays of the neutron, pions,  kaons etc., 
as well as to the interaction 
 \be{muon}
- \frac{4G_F}{\sqrt2}\, \big(\ov{e_L} \gamma_\mu \nu_{e} \big) \big(\ov{\nu_{\mu}} \gamma^\mu \mu_L \big) 
\ee
responsible for the muon decay.  
All these couplings contain the Fermi constant  $G_F/\sqrt2=g^2/8M_W^2$.  

Precision experimental data on kaon decays, in combination with the lattice QCD calculations  
of the decay constants and form-factors, provide accurate information about  $\vert V_{us}\vert $. 
On the other hand,  recent calculations of short-distance radiative corrections 
in the neutron decay allow to determine $\vert V_{ud}\vert $ with a remarkable precision. 

In this paper we analyze the present individual determinations 
of $V_{ud}$ and $V_{us}$ and find significant (more than $4\sigma$) deviation 
from the CKM unitarity (\ref{unitarity}).      
We discuss two possible BSM scenarios which can explain this deviation. 
In the first one the three-family unitarity is extended to four species, 
by introducing the 4th down-type quark $b'$ with mass of few TeV.  
The second scenario assumes the 
existence of horizontal gauge symmetry between the lepton families 
which is spontaneously broken at the scale of few TeV.   
The corresponding flavor changing gauge bosons 
 induce the effective  four-lepton interaction 
having exactly the same form as (\ref{muon}), with the new Fermi-like constant $G_\cF$.    
In this case, muon lifetime would determine $G_\mu = G_F+G_\cF$ rather than  $G_F$.  
In this way,  one can nicely restore the three family 
unitarity (\ref{unitarity}) without introducing new quark species.  
We discuss implications of these scenarios for the lepton flavor violation (LFV) 
and for the Standard Model precision tests. 
At the end, we also discuss the problem of neutron lifetime   
related to the discrepancy between its values measured  
using two different (trap and beam) methods.

\medskip 
\noindent {\bf 2.}
The most precise determination of $\vert V_{ud} \vert $  is obtained 
from superallowed $0^+\!-0^+$ nuclear $\beta$-decays which are pure Fermi transitions 
sensitive only to the vector coupling constant $G_V=G_F \vert V_{ud} \vert$ \cite{Hardy}: 
\be{Vud-super}
\vert V_{ud} \vert^2 = \frac{K }{ 2 G_F^2 \cF t\, (1+ \Delta_R^V)}  
= \frac{0.97147(20)}{ 1+ \Delta_R^V } 
\ee
where $K = 2\pi^3 \ln 2/m_e^5 = 8120.2776(9) \times 10^{-10}$~s/GeV$^4$ 
and $\cF t$ is the nucleus independent value obtained from the individual 
$ft$-values of different $0^+\!-0^+$ nuclear transitions  
by absorbing  in the latter all nucleus-dependent corrections, 
while $ \Delta_R^V$ accounts for  short-distance (transition independent)
radiative corrections. 
For the second step, we  take  $\cF t = 3072.07(72)$~s  \cite{Hardy2}    
obtained  by averaging the individual  $\cF t$-values 
for fourteen superallowed $0^+\!-0^+$  transitions determined with 
 the best experimental accuracy,    
and plug in the Fermi constant as $G_F=G_\mu  = 1.1663787(6) \times 10^{-5}$~GeV$^{-2}$ 
determined from the muon decay \cite{mulan}.  
The major uncertainty is related to the 
so called inner  
radiative correction $\Delta_R^V$. 


The element $\vert V_{us} \vert $ can be determined from analysis of 
semileptonic $K\ell3$ decays ($K_L \mu3$, $K_L e3$,  $K^{\pm}e3$, etc.)
\cite{Moulson}: 
\be{f+}
f_+(0) \vert V_{us}\vert   = 0.21654\pm 0.00041 
\ee
where $f_+(0)$  is the $K\to \pi \ell \nu$ vector form--factor at zero momentum transfer. 
On the other hand, by comparing the kaon and pion inclusive radiative 
decay rates $K\to \mu\nu(\gamma)$ and $\pi \to \mu\nu(\gamma)$, one obtains \cite{PDG2018}:
\be{fK}
\vert V_{us}/V_{ud} \vert \times (f_{K^\pm}/f_{\pi^\pm}) = 0.27599 \pm 0.00038 \, . 
\ee
Hence,  the values  $\vert V_{us}\vert$ and  $\vert V_{us}/V_{ud} \vert$ can be 
independently determined using the lattice QCD results for the form--factor 
$f_+(0)$ and the decay constant ratio $f_K/f_\pi$.  

\medskip 
\noindent
{\bf 3.} 
Let us first consider the values of the CKM matrix elements 
$\vert V_{us} \vert $, $\vert V_{ud} \vert $ and their ratio $\vert V_{us}/V_{ud} \vert $ 
as quoted  by Particle Data Group (PDG) review 2018 \cite{PDG2018}:  
\begin{eqnarray}\label{data-PDG}
&&   \quad  \vert V_{us} \vert = 0.2238(8)     \nonumber \\ 
&&    \vert V_{us}/V_{ud} \vert = 0.2315(10)      \\
&&   \quad \vert V_{ud} \vert = 0.97420(21)   \nonumber 
\end{eqnarray} 
Here $\vert V_{us} \vert$ and $ \vert V_{us}/V_{ud} \vert $ 
are obtained respectively  from Eqs. (\ref{f+}) and (\ref{fK})  using the FLAG 2017 averages 
of 3--flavor lattice QCD simulations $f_+(0) = 0.9677(27)$ and $f_{K^\pm}/f_{\pi^\pm}=1.192(5)$  
 \cite{FLAG2017}. $\vert V_{ud} \vert$  is obtained from Eq.~(\ref{Vud-super}) 
by taking  $\Delta_R^V = 0.02361(38)$ as calculated in Ref. \cite{Marciano:2006}.

 \begin{figure}
\includegraphics[width=0.38\textwidth]{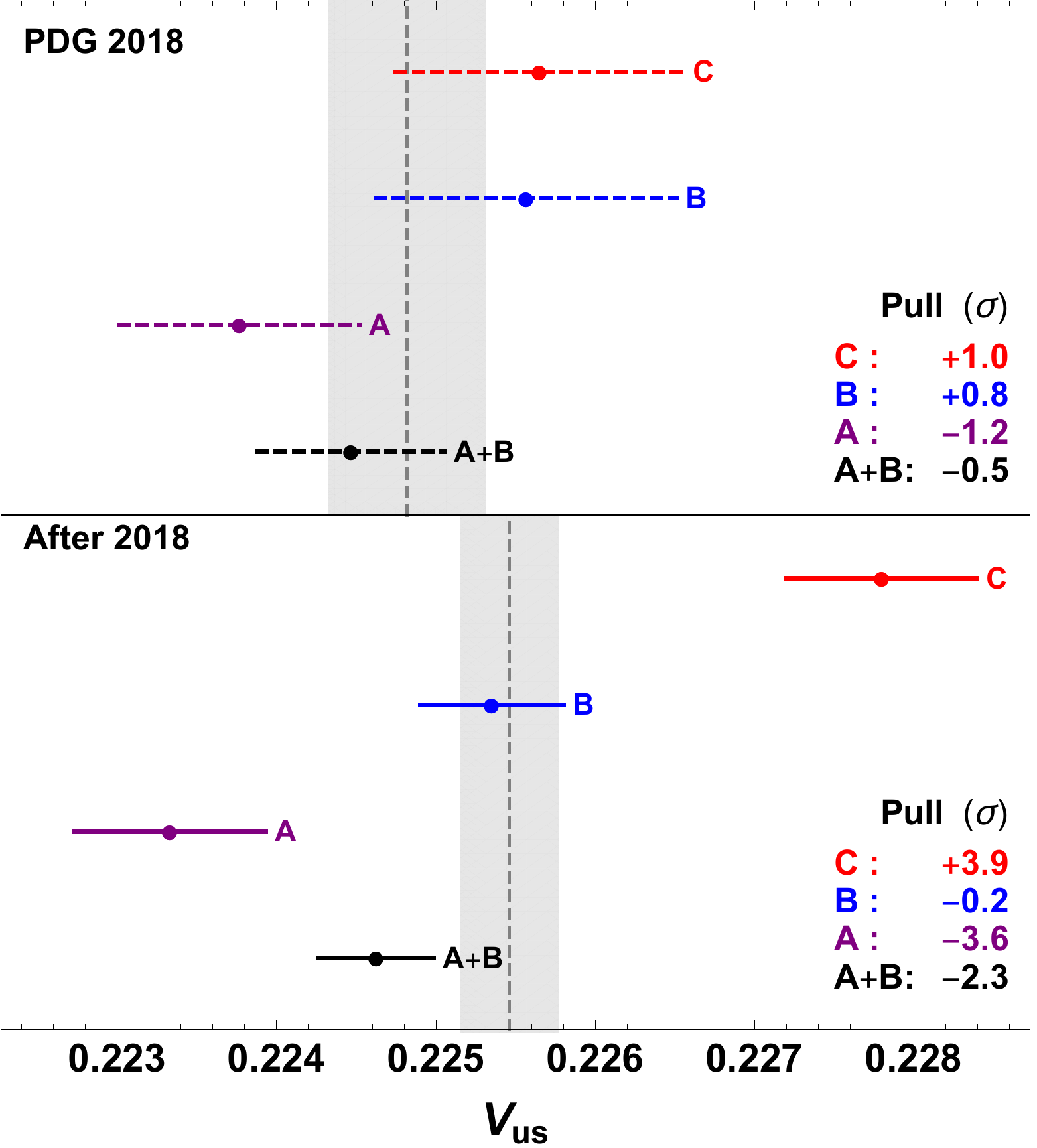}
\caption{{\it Upper panel:}  Three independent $\vert V_{us} \vert$ determinations A, B, C 
obtained from the PDG 2018 data (\ref{data-PDG}) by assuming the CKM unitarity. 
The grey shaded band corresponds to the average 
 A+B+C (with formal error not rescaled by a factor $\sqrt{\chi^2_{\rm dof}} $). 
 Pulls of C, B, A  and A+B are shown.  
{\it Lower panel:}  The same for the  A, B, C values obtained from new data (\ref{data-new}).  
  }
\label{fig-1}
\end{figure}

By imposing the CKM unitarity (\ref{unitarity}), the three data  (\ref{data-PDG}) 
reduce to three independent determinations of $\vert V_{us} \vert$. 
These determinations     
shown as A, B, C  in upper panel of Fig.~\ref{fig-1} 
(see also Table~\ref{Table} for numerical values) 
are compatible within their error-bars.\footnote{Throughout this paper A is the direct determination  
of $\vert V_{us} \vert$  obtained from Eq. (\ref{f+}).  B and C are the values of $\vert V_{us} \vert$
 obtained respectively from $ \vert V_{us}/V_{ud} \vert$  and  $\vert V_{ud} \vert$ 
by assuming unitarity. 
 } 
Namely, B and C are almost  equal while there is a modest tension  ($1.4\sigma$)  
between A and B.  Their average $\overline{A\!+\!B} =0.2245(6)$, practically 
coincides with the PDG 2018 average of $\vert V_{us} \vert$ \cite{PDG2018}.  
By averaging all three values we get  
 $\overline{A\!+\!B\!+\!C}= 0.2248(5)$ 
 with $\chi^2_{\rm dof} =1.7$. 
Pulls of  A, B and C relative to this average  (given in Fig.~\ref{fig-1})
are compatible with a standard deviation. 
Summarizing, the dataset  (\ref{data-PDG}) adopted from PDG 2018 \cite{PDG2018}
is  consistent with the CKM unitarity (\ref{unitarity}).

However,  recent  progress  in the determination of the CKM elements  
allows to test the unitarity with improved precision.  
Significant  redetermination of $\vert V_{ud} \vert$ 
is related  to new calculation of inner radiative corrections with reduced hadronic uncertainties,  
 $\Delta_R^V = 0.02467(22)$ \cite{Seng:2018}. 
Employing  also the recent result $f_+(0) = 0.9696(18)$ 
from new 4--flavor $(N_f=2\!+\!1\!+\!1)$ lattice QCD simulations 
\cite{Bazavov}  and the FLAG 2019 four-flavor average   $f_{K^\pm}/f_{\pi^\pm}=1.1932(19)$  
 \cite{FLAG2019}, one arrives to the following set:\footnote{
Alternatively, one could use  the FLAG 2019 average $f_+(0)=0.9706(27)$ \cite{FLAG2019}
(not including result  of Ref. \cite{Bazavov}) leading to a minor change of  
$\vert V_{us}\vert $ in (\ref{data-new})  from $0.22333(60)$ to $0.22310(75)$. 
}
\begin{eqnarray} \label{data-new}
&&  \quad  \vert V_{us} \vert = 0.22333(60)     \nonumber \\ 
&&  \vert V_{us}/V_{ud} \vert = 0.23130(50)      \\
&&   \quad \vert V_{ud} \vert = 0.97370(14) \nonumber    
\end{eqnarray} 
This dataset, again by imposing the CKM unitarity, reduces to 
independent $\vert V_{us} \vert$ values A, B, C shown in lower panel of Fig.~\ref{fig-1}  
(numerical values are given in  Table~\ref{Table}). 
%

Now we see that the values A, B, C 
are in tensions among each other. 
Namely, there is a $5.3\sigma$ discrepancy  between A and C,  and $3.2\sigma$ 
between B and C. The tension between the determinations A and B, both from kaon physics, 
is $2.7\sigma$. More conservatively, one can take 
their average $\overline{A\!+\!B}$.  The discrepancy of the latter with C is $4.5\sigma$. 
Fitting these values, we get  
$\ov{A\!+\!B\!+\!C} = 0.22546(31)$ but  the fit is bad, $\chi^2_{\rm dof} =13.9$. 
C, A and A+B have large pulls, 
$3.9\sigma$, $-3.6\sigma$ and $-2.3\sigma$. 

\begin{figure}
\includegraphics[width=0.45\textwidth]{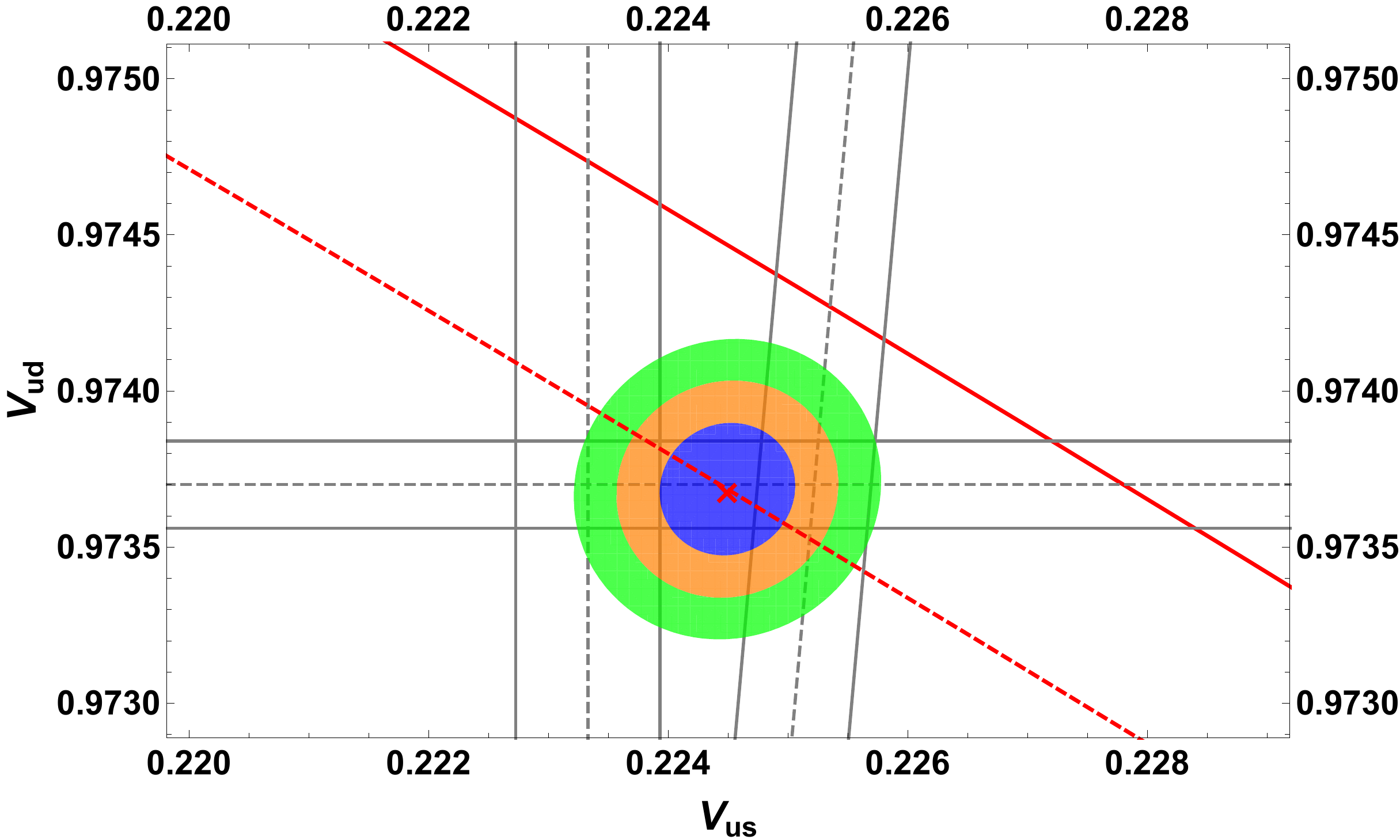}
\caption{The horizontal, vertical and slightly bended bands correspond  
to $\vert V_{ud} \vert $, $\vert V_{us} \vert $ and $\vert V_{us}/V_{ud} \vert $
from (\ref{data-new}).
The best fit point (red cross) and  $1$, $2$ and $3\sigma$  contours are shown. 
The red solid line corresponds to the three family unitarity condition (\ref{unitarity}),  
and the dashed red line corresponds to the "extended"  unitarity (\ref{CKM-mod}) 
with $\vert V_{ub'}\vert = 0.04$. }
\label{fig-2CKM}
\end{figure}

This tension can be manifested also by analyzing  the data (\ref{data-new}) in 
a different way.  Without imposing the unitarity condition (\ref{unitarity}), 
we perform a two parameter fit of the three independent values (\ref{data-new}). 
In Fig.~\ref{fig-2CKM}  we show the gaussian hill of the probability distribution with 
the confidence level (C.L.)  contours around the best fit point    
($\vert V_{us} \vert = 0.22449$, $\vert V_{ud} \vert = 0.97369$),     
with  $\chi^2_{\rm min} = 6.1$.  (This $\chi^2$--value seems  large for a two parameter fit, 
but it is dominated by the tension between the determinations A and B  of 
$\vert V_{us}\vert$ from the kaon data and perhaps 
this tension will disappear  with more accurate  lattice simulations.)
The red solid line corresponding to the three family unitarity condition 
$\vert V_{ud} \vert^2 + \vert V_{us} \vert^2 = 1- \vert V_{ub} \vert^2 = 1- O(10^{-5})$    
is about $4.3\sigma$ away  from this hill  ($\Delta\chi^2 = 21.6$).  
In other words,  the new (after 2018) dataset (\ref{data-new}) 
disfavors the CKM unitarity at $99.998 \%$ C.L.  

\begin{figure}
\includegraphics[width=0.38\textwidth]{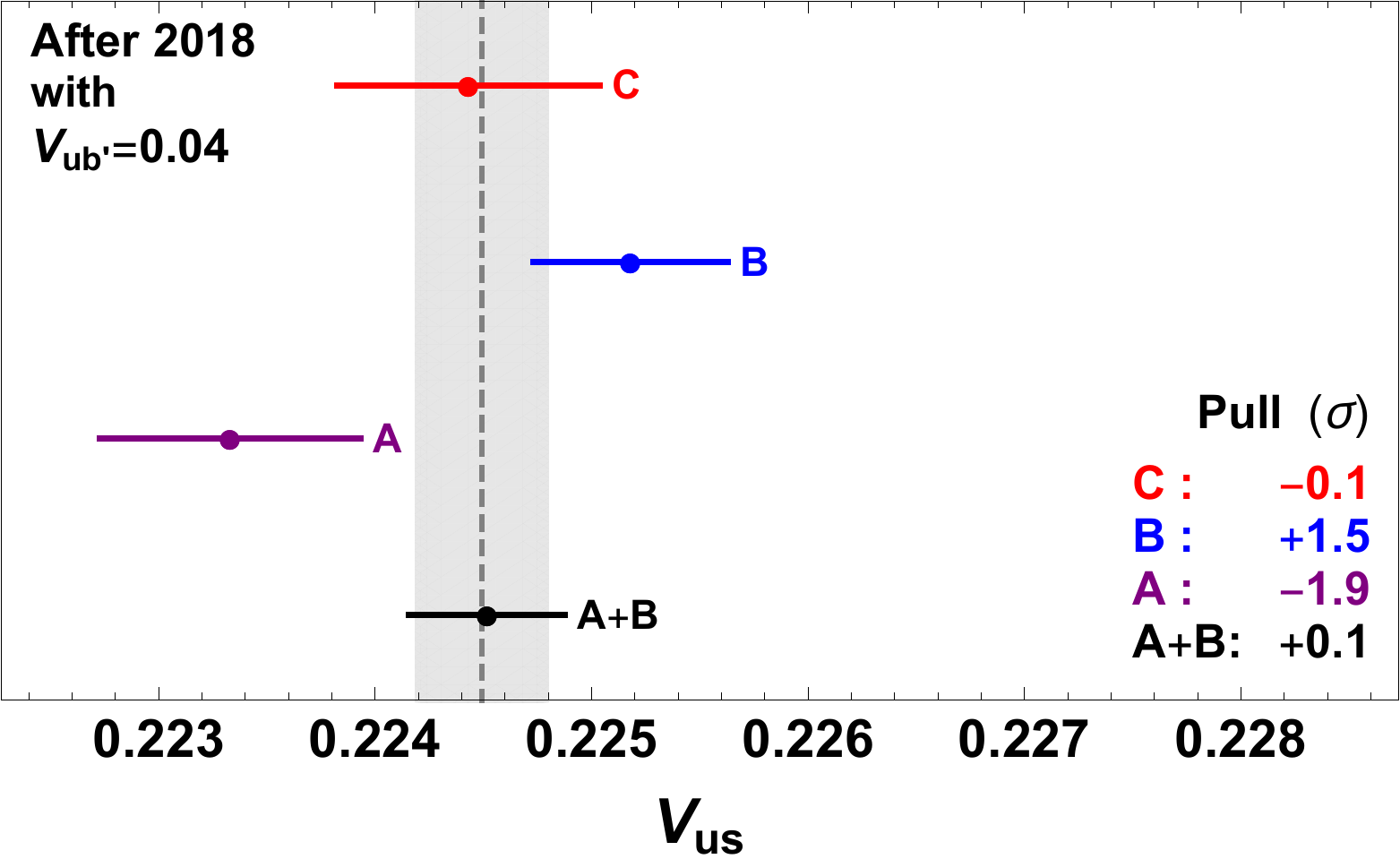}
\caption{ Determinations of $\vert V_{us} \vert$  
obtained  from the dataset (\ref{data-new}) 
using Eq. (\ref{CKM-mod}) with $\vert V_{ub'} \vert = 0.04$. 
 }
\label{fig-3}
\end{figure}

\medskip 
\noindent
{\bf 4.} {\it ``If the Hill will not come to the CKM, the CKM will go to the Hill."} 
The unitarity line 
can be moved down towards the probability distribution hill  in Fig.~\ref{fig-2CKM} 
if the unitarity condition is extended to more families.
One can introduce, besides the three down quarks $d,s,b$, 
a 4-th state $b'$ which is also involved in  quark mixing. 
Then the first row unitarity condition will be  modified to 
\be{CKM-mod} 
\vert V_{ud} \vert^2 + \vert V_{us} \vert^2 +\vert V_{ub} \vert^2  + \vert V_{ub'} \vert^2 = 1 \, .
\ee
In particular, the red dashed line in Fig.~\ref{fig-2CKM} passing through the best fit point on 
the top of the probability hill corresponds to $\vert V_{ub'} \vert = 0.04$ 
(at $95$~\% C.L. this additional mixing  is limited as $\vert V_{ub'} \vert = 0.04 \pm 0.01$). 
Plugging this value in Eq.~(\ref{CKM-mod}), 
the dataset (\ref{data-new}) gives  the modified determinations of $\vert V_{us} \vert$ 
for the three cases named above as A, B and C 
(for numerical values see  in 3rd column of Table~\ref{Table}). 
Clearly, the case A in this list remains the same as in 2nd column but B and especially 
C are shifted down. 
Fig.~\ref{fig-3} shows that consistency between these values is significantly improved 
compared to lower panel of Fig.~\ref{fig-1}.  
The fit  for $\ov{A\!+\!B\!+\!C}$  is acceptable, $\chi^2_{\rm dof} =3$. 
Pulls of C and A+B are practically vanishing. There remains a tension 
between A and B but it is softened to $2.4\sigma$ from $2.7\sigma$ of Fig.~\ref{fig-1}.


Let us discuss now in which conditions one could obtain so large mixing 
with the 4th species, $\vert V_{ub'} \vert \approx 0.04$. 
 In the SM the three families ($i=1,2,3$ is the family index) of 
left-handed (LH) quarks $Q_{Li}=(u_i,d_i)_L$ and leptons $\ell_{Li}=(\nu_i,e_i)_L$
transform as weak isodoublets 
of $SU(2)\times U(1)$ and the right-handed (RH) quarks $u_{Ri},d_{Ri}$ and leptons $e_{Ri}$ 
are the isosinglets.  
The existence of a fourth sequential family is excluded by the SM precision tests 
in combination with the direct limits from the LHC,  
but one can introduce additional vector-like fermions. 
Let us briefly sketch a vanilla picture of this type, 
adding just a vector-like couple of isosinglet down-type quarks $b'_L,b'_R$ having 
a large Dirac mass $M \ov{b'} b'$, $b'=b'_L+b'_R$.
In this way,  we obtain the modified $3\times 4$  matrix of the quark mixing in left-handed 
charged current: 
\be{CKM4} 
 \tilde{V}_{CKM}=\left(\begin{array}{cccc}
V_{ud}&V_{us}&V_{ub} & V_{ub'}\\
V_{cd}&V_{cs}&V_{cb} & V_{cb'}\\
V_{td}&V_{ts}&V_{tb} & V_{tb'} \, .
\end{array}\right)
\ee 
The condition (\ref{CKM-mod}) regards the first row of this matrix.\footnote{One 
can introduce 
also a fourth upper quark $t'$, so that two singles $b'$ and $t'$ would form a family in some sense, 
and the mixing matrix (\ref{CKM4}) would become a $4\times 4$ matrix.   
However,  this modification of the minimal picture is irrelevant since 
$t'$ will have no impact on the first row unitarity. In addition, it can be easily shown  
that introduction of fourth vector-like isodoublet family $Q'_{L,R} = (t',b')_{L,R}$ 
cannot  generate large enough mixing $ V_{ub'} $.  
}   


Without losing generality, the Yukawa terms  can be 
 divided in two parts. The first part
%
 \be{Yukawa-SM}
Y_u^{ij} \tilde\phi \, \ov{Q_{Li}} u_{Rj}  +   Y_d^{ij} \phi  \, \ov{Q_{Li}} d_{Rj}  
+ Y_e^{ij} \, \phi \, \ov{\ell_{Li}} e_{Rj}  
\, + \, {\rm h.c.} \,  
\ee
comprises the SM Yukawa terms of  three standard families with the Higgs doublet $\phi$, 
$Y_{u,d,e}$ being the Yukawa constant matrices and $\tilde\phi=i\tau_2 \phi^\ast$. The second part 
\be{Yukawa-BSM}
h_i \phi\, \ov{Q_{Li}} b'_R + M \,\ov{b'_L} b'_R  \, + \, {\rm h.c.}   
\ee 
involves the extra state $b'$.
Fermion masses emerge from the  vacuum expectation value (VEV) of the Higgs, 
 $\langle \phi^0\rangle = v_{\rm w} = 174$ GeV  
(for a convenience,  we use this normalization of the Higgs  VEV 
instead of  ``standard" normalization $\langle \phi\rangle = v/\sqrt2$, i.e. $v=\sqrt2 v_{\rm w}$).  

Without loss of generality, the matrix $Y_d$ can be chosen diagonal,
$Y_d = Y_d^{\rm diag}= {\rm diag}(y_d,y_s,y_b)$. 
The Yukawa terms in (\ref{Yukawa-BSM}) 
induce the mixing of three known quarks $d,s,b$  to the 4th quark $b'$.
 Thus,  $4\times 4$ mass matrix of all down-type quarks has a form: 
\be{M4} 
\left(\begin{array}{cccc}
Y_d v_{\rm w} & 0 & 0  & h_d v_{\rm w} \\
0 & Y_s v_{\rm w} & 0  & h_s v_{\rm w} \\
0 & 0 & Y_b v_{\rm w} & h_b v_{\rm w}  \\
0 & 0 &   0   &  M
\end{array}\right)
\ee 
In this basis, 
the up quark Yukawa matrix is non-diagonal, 
$Y_u = V_{uL} Y_u^{\rm diag} V_{uR}^\dagger$,  $Y_u^{\rm diag}=  {\rm diag}(y_u,y_c,y_t) $, 
where $3\times 3$  unitary matrix $V_{uL}$ in fact  determines the ordinary three-family part (\ref{CKM})
of the quark mixing, i.e. $V_{Lu}^\dagger = V_{\rm CKM}$.  
Then  $3\times 4$ extended mixing matrix 
$\tilde{V}_{\rm CKM}$ (\ref{CKM4}) is completed by diagonalization of the matrix (\ref{M4}). 
Namely, the off-diagonal terms  in (\ref{M4}) determine the mixing of $d,s,b$ with $b'$,  
 $V_{ub'} \simeq h_d v_{\rm w}/M$,   $V_{cb'} \simeq h_s v_{\rm w}/M$ and  
$V_{tb'} \simeq h_bv_{\rm w}/M$ which values are generically complex.  
This mixing practically does not affect the diagonal elements 
in (\ref{M4}). Hence, $m_{d,s,b} = Y_{d,s,b} v_{\rm w}$ and $m_{b'} =M$. 

The LHC limit on extra $b'$ mass  $M> 880$~GeV \cite{PDG2018} implies that 
$\vert V_{ub'}\vert  \simeq  0.04$ 
can be obtained if $h_d > 0.2$, much larger than the Yukawa constant $Y_b$. 
 In turn,  by taking $\vert V_{ub'} \vert  > 0.03$ in  $M = h_dv_{\rm w}/\vert V_{ub'}\vert$, 
 and assuming (for the perturbativity) $h_d < Y_t \simeq 1$, 
 we get an upper limit on the extra quark mass, $M < 6$~TeV or so. 
 

\begin{figure}[t]
\includegraphics[width=0.2\textwidth]{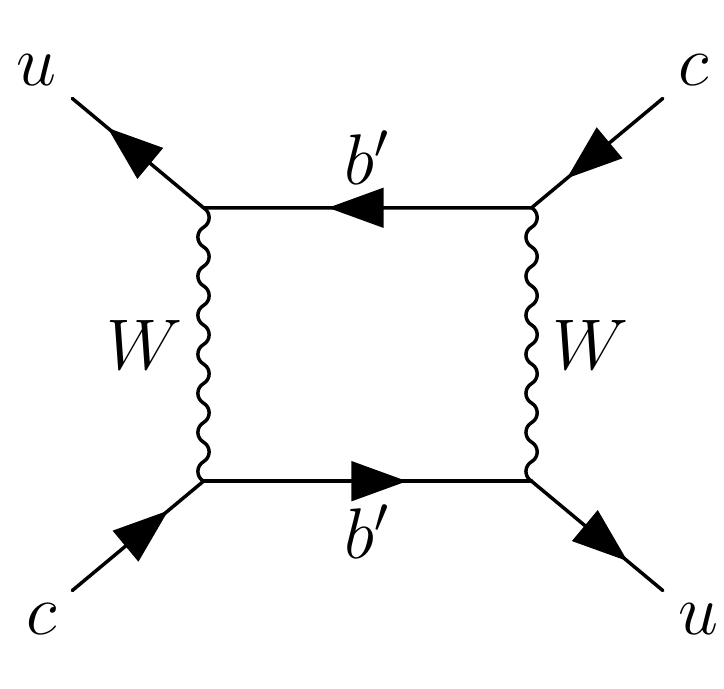} 
\hspace{7mm}
\includegraphics[width=0.2\textwidth]{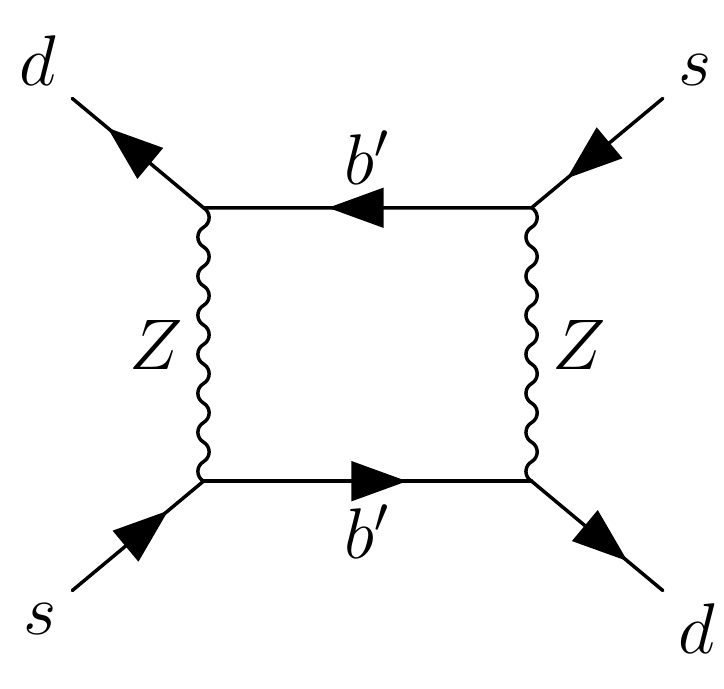}
\caption{$Wb'$ box inducing $D^0\!-\!\bar{D}^0$ mixing and $Zb'$ box inducing $K^0\!-\!\bar{K}^0$ mixing.  }
\label{diagram-KK}
\end{figure}

The extension of the SM by adding an extra isosinglet quark $b'$ generates significant contributions 
in flavor-changing processes. E.g., $Wb'$ box diagram shown in Fig.~\ref{diagram-KK} 
induces $D^0\!-\!\bar{D}^0$ mixing. For $\vert V_{ub'} \vert \simeq 0.04$,  
its contribution would exceed the experimental value of 
their mass splitting, $\Delta M \simeq 10^{10}$~s$^{-1}$, unless 
$\vert V_{cb'}/V_{ub'} \vert  \times (M/1~{\rm TeV})  < 1/3$ or so. 
In addition, as one can see from (\ref{Yukawa-BSM}),  
the 4th quark has tree level flavor-changing couplings with the Higgs boson $H$   
and also with $Z$-boson: 
\beqn{FCNC-Z} 
&&  \frac{M}{\sqrt2 v_{\rm w}} H 
\big(  V_{ub'}  \ov{d_L} + V_{cb'} \ov{s_L} + V_{tb'} \ov{b_L}\big)\, b'_R 
 +{\rm h.c.}   \nonumber \\ 
&& \frac{g }{2 c_W}  Z_\mu 
\big( V_{ub'} \ov{d_L}  + V_{cb'} \ov{s_L} + V_{tb'} \ov{b_L}  \big) \gamma^\mu b'_L 
+ {\rm h.c.} 
\eeqn
Then the $Zb'$ box diagram shown in Fig.~\ref{diagram-KK}  contributes to $K^0\!-\!\bar{K}^0$ mixing.  
Interestingly, for $\vert V_{ub'} \vert \simeq 0.04$ 
this new contribution in CP-violating $\epsilon_K$--parameter would be larger 
than the SM one unless 
$\arg(V_{cb'}/V_{ub'}) \times \vert V_{cb'}/V_{ub'} \vert  \times (M/1~{\rm TeV})  < 1/10$ or so. 
For $\vert V_{tb'} \vert \sim \vert V_{ub'} \vert =0.04$, the analogous $Zb'$ box diagram with external 
$b$ quark would give a contribution to $B_d-\bar{B}_d$ meson mixing comparable to the SM contribution.

These flavor-changing and CP-violating effects can be suppressed if $ V_{cb'} $ and $ V_{tb'} $ 
 are much less than $V_{ub'}$, or at least have rather small complex parts. 
 (Accidentally, $\vert V_{ub'} \vert \simeq 0.04 $ is comparable to $\vert V_{cb} \vert$ 
and ten times larger than $\vert V_{ub} \vert$.) 
The picture with the 4th state $b'$ having a larger mixing  with the first   
family  than with (heavier) 2nd and 3rd families  
looks somewhat {\it ad hoc}, but it is not excluded 
by the present experimental limits. 
The implications of a TeV scale extra quark $b'$ 
with significant $\vert V_{ub'}\vert $ mixing  
deserve careful analysis.


\medskip 
\noindent
{\bf 5.} {\it ``But what if the Hill comes to the CKM?"}~ 
Here we discuss just the opposite possibility: instead of moving the unitarity line 
 to the probability distribution Hill in Fig.~\ref{fig-2CKM}, 
we move  the Hill towards the unitarity line. 
 
 \begin{figure}[t]
\includegraphics[width=0.23\textwidth]{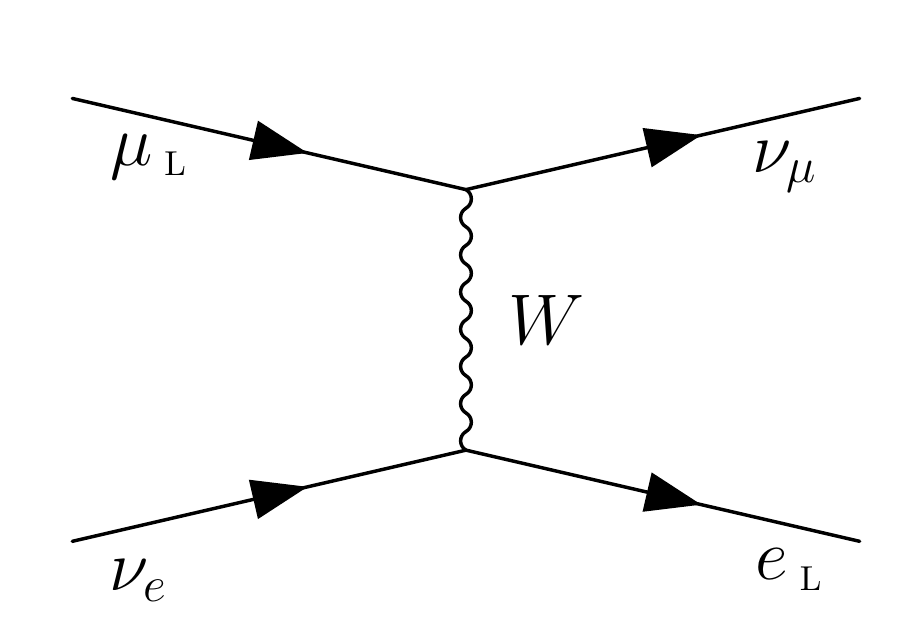} \hspace{2mm}
\includegraphics[width=0.23\textwidth]{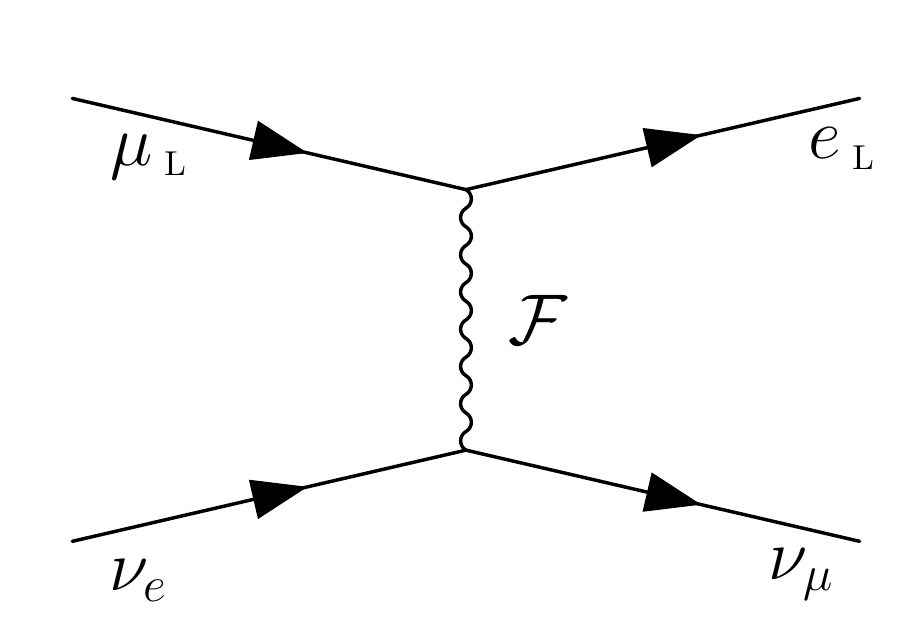}
\caption{The SM contribution to the muon decay mediated by $W$-boson (left), 
and the BSM contribution mediated by the flavor-changing $\cF$--boson  (right).  } 
\label{diagrams}
\end{figure}
 
\begin{figure}
\includegraphics[width=0.45\textwidth]{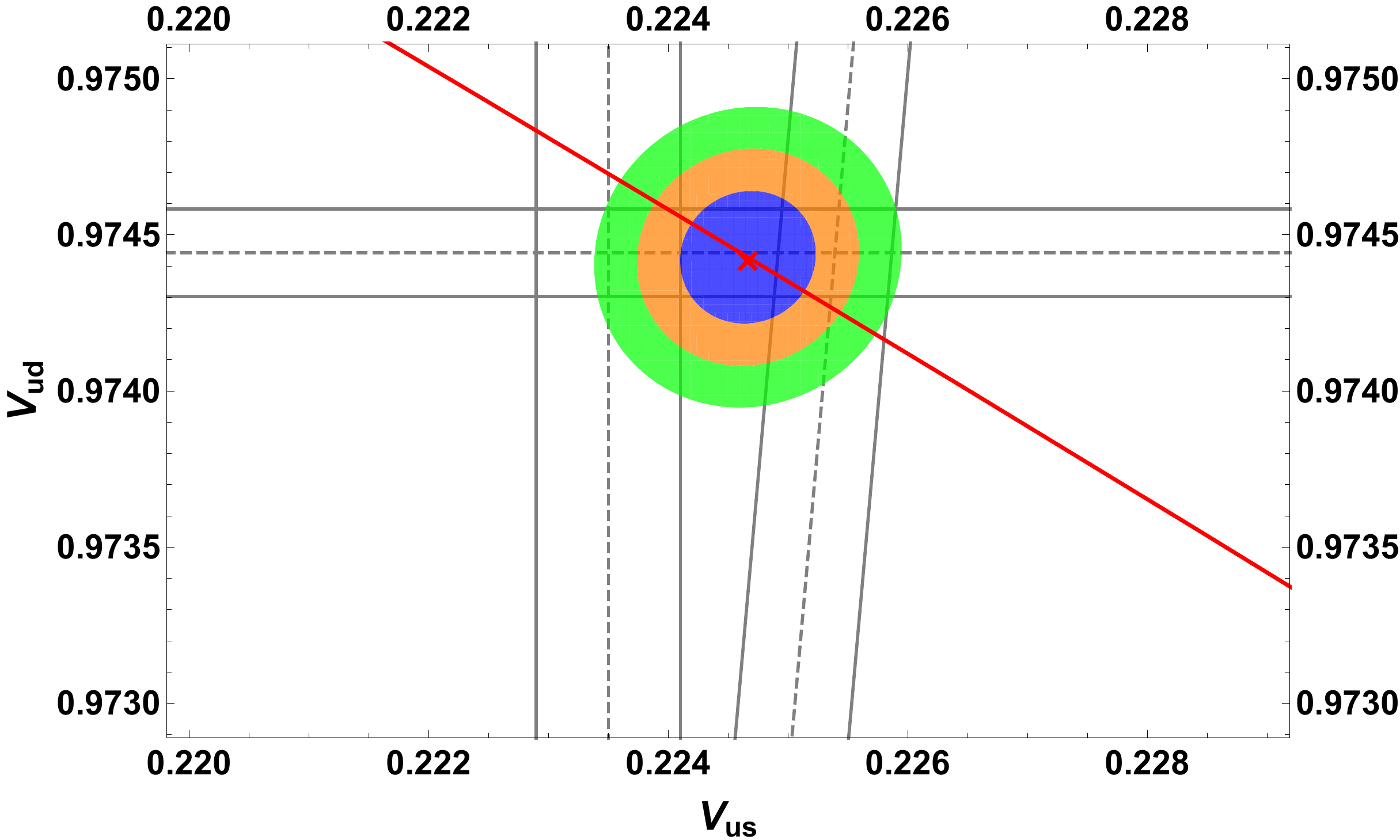}
\caption{
The same as on Fig.~\ref{fig-2CKM} but with the bands of $\vert V_{ud}\vert$ 
and $\vert V_{us}\vert$ 
rescaled up by a factor $1+\delta_\mu=1.00076$ 
while the band for $\vert V_{us}/V_{ud}\vert$ remains the same.  
The red line corresponds to three-family unitarity (\ref{unitarity}) as in Fig.~\ref{fig-2CKM}.
}
\label{fig-2H}
\end{figure}

Namely, we consider that  the Fermi constant $G_F$ in the effective interaction 
(\ref{ud}) which is responsible for leptonic decays of hadrons 
can be different from the effective constant $G_\mu$ determined from the muon lifetime. 
We assume that besides the SM interaction (\ref{muon}) mediated by charged $W$--boson, 
there is also  a new operator 
\be{new}
-\frac{4G_\cF}{\sqrt2} (\ov{e_L} \gamma_\mu \mu_L)(\ov{\nu_\mu} \gamma^\mu \nu_e)
\ee 
mediated by a hypothetical lepton flavor changing neutral gauge boson $\cF$.   
The respective diagrams, shown in Fig.~\ref{diagrams},  have positive interference  
for the muon decay. 
Namely,  by Fierz transformation  this new operator can be brought to the form (\ref{muon}),   
so that the sum of these two diagrams effectively gives the operator 
 \be{muon1}
- \frac{4G_\mu}{\sqrt2}\, \big(\ov{e_L} \gamma_\mu \nu_{e} \big) \big(\ov{\nu_{\mu}} \gamma^\mu \mu_L \big) \, , 
\ee
the same as (\ref{muon}) but with the coupling constant 
\be{Gmu}
G_\mu = G_F + G_\cF=G_F(1+ \delta_\mu), \quad\quad 
\frac{G_\cF}{G_F} \equiv \delta_\mu >0 \, . 
\ee
Constant $G_\mu=1.1663787(6) \times 10^{-5}$~GeV$^{-2}$ is 
determined with great precision from the muon decay \cite{mulan}.    
Now Eqs. (\ref{Vud-super}) and (\ref{f+}),  instead of $\vert V_{ud}\vert $ and $\vert V_{us}\vert $, 
are determining respectively the values $\vert V_{ud} \vert \times G_F/G_\mu $ and 
$\vert V_{us} \vert \times G_F/G_\mu $.  
Instead the value of $\vert V_{us}/V_{ud}\vert$ determined from (\ref{fK}) remains unchanged 
since the Fermi constant cancels out. 
 Thus, under our hypothesis, the dataset (\ref{data-new}) should be modified to the following:
\begin{eqnarray} \label{data-new-delta}
&&  \vert V_{us} \vert = 0.22333(60) \times (1+\delta_\mu)   \nonumber \\ 
&&  \vert V_{us}/V_{ud} \vert = 0.23130(50)      \\
&&   \vert V_{ud} \vert = 0.97370(14) \times (1+\delta_\mu)  \nonumber    
\end{eqnarray} 
 Now, involving the extra parameter $\delta_\mu$ but  assuming the 3-family unitarity (\ref{unitarity}),  
the fit of the above dataset has acceptable quality, $\chi^2 = 6.1$,   and the best fit point 
corresponds to $\delta_\mu=0.00076$. 
This situation is shown in Fig.~\ref{fig-2H} in which 
the values of $\vert V_{ud} \vert$ and $\vert V_{us} \vert$  are determined by taking 
$\delta_\mu=0.00076$. 
By this choice of the extra parameter  
 the
 fit becomes perfectly compatible with the unitarity (\ref{unitarity}).  
The probability distribution Hill is moved up so that its top now lies on the unitarity line.  

\begin{figure}
\includegraphics[width=0.4\textwidth]{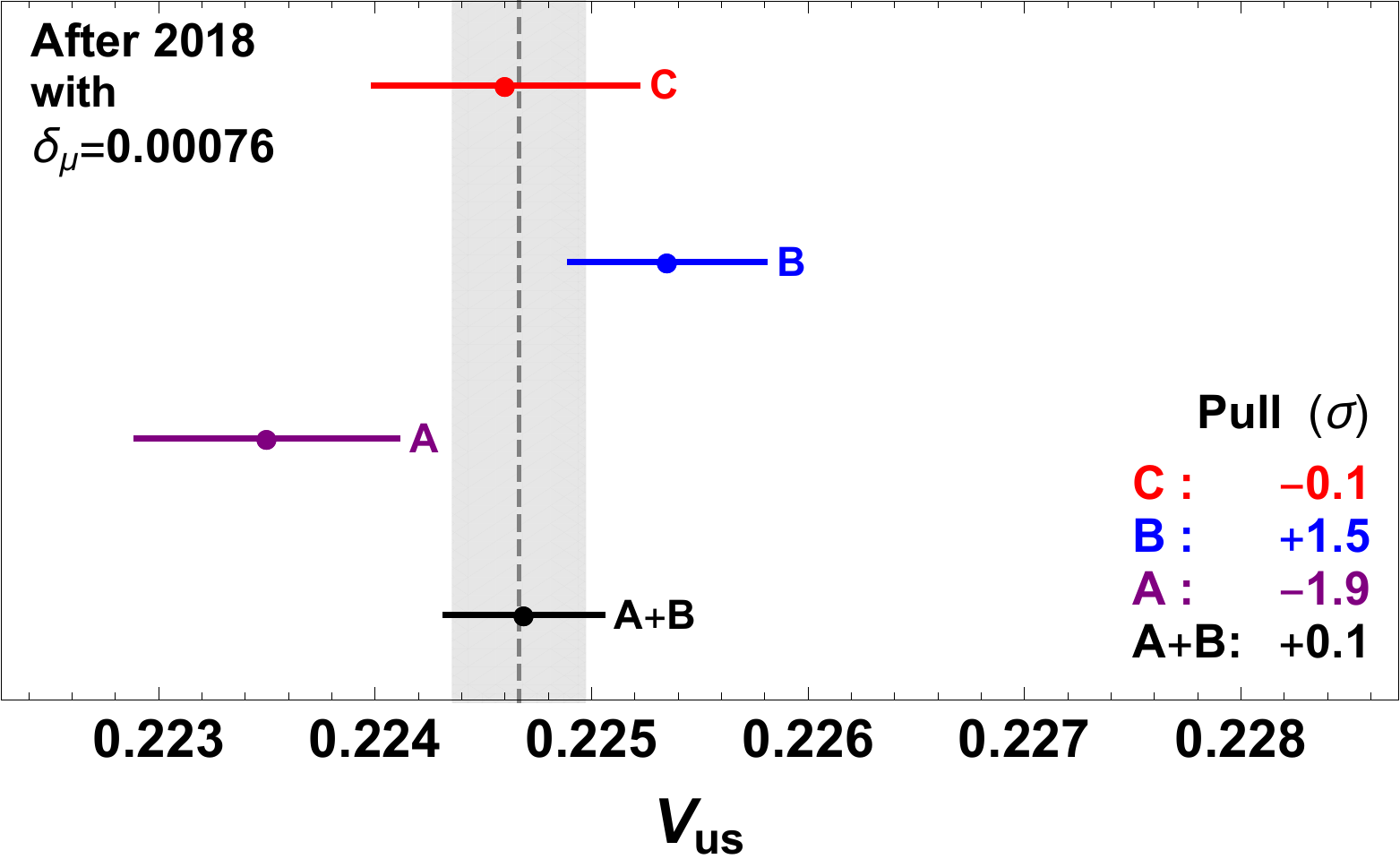}
\caption{Determinations of $\vert V_{us}\vert$ obtained from (\ref{data-new-delta}).  }
\label{fig-1H}
\end{figure}

By imposing the unitarity condition 
$\vert V_{ud} \vert^2 + \vert V_{us} \vert^2 = 1- \vert V_{ub} \vert^2$, 
the list (\ref{data-new-delta}) can be transformed in $\delta_\mu$ dependent 
determinations A, B, C of $\vert V_{us} \vert$. 
Fig.~\ref{fig-1H} shows these determinations for $\delta_\mu=0.00076$. 
Taking into account that $G_F/\sqrt2 = g^2/8M_W^2 = 1/4v_{\rm w}^2$, where 
$v_{\rm w} = 174$~GeV is the weak scale,  and parametrizing similarly $G_\cF/\sqrt2 = 1/4v_\cF^2$, 
we see that $\delta_\mu = G_\cF/G_F = 0.00076$ corresponds to $v_\cF/v_{\rm w} = 36.3$, or 
to the flavor symmetry breaking scale $v_\cF = 6.3$ TeV.  More widely, the range of 
$\delta_\mu$ consistent with unitarity at the 68\% C.L. 
is $\delta_\mu = (7.6 \pm 1.6) \times 10^{-4}$ which corresponds to the new scale 
in the interval $v_\cF = [5.7 \div 7.1]$~TeV.

\medskip

\noindent {\bf 6.} 
The non-abelian gauge horizontal flavor symmetry $G_H$ between the fermion families can be the key 
for understanding the quark and lepton mass and mixing pattern \cite{PLB83,Chkareuli}. 
Namely, the form of the Yukawa matrices $Y_{u,d,e}$ in 
(\ref{Yukawa-SM}) can be determined by the $G_H$ symmetry breaking pattern, i.e. by 
the VEV structure of the horizontal scalar fields (flavons) 
responsible for this  breaking. Then the fermion mass hierarchy is related to 
the hierarchy between these VEVs. 
In Refs. \cite{PLB83} this conjecture 
was coined as {\it hypothesis of horizontal hierarchies}.  
In this picture the fermion masses emerge from the higher order operators 
involving, besides the Higgs doublet $\phi$, also flavon scalars which transfer their VEV structure 
to the Yukawa matrices $Y_{u,d,e}$. These so called ``projective" operators 
in the UV-complete renormalizable theory can be obtained via integrating out some extra heavy 
fields, scalars \cite{Chkareuli} or vector-like fermions \cite{PLB83}.   
In particular, this concept implies that the fermion 
masses cannot emerge if $G_H$ symmetry  is unbroken. 
Thus,  $G_H$ cannot be a vector-like symmetry  but it should have a chiral character 
transforming the LH and RH particle species in different representations. 
In particular, in Refs. \cite{Chkareuli,PLB83,Khlopov,SO10} the horizontal symmetry $G_H$ 
 was considered as $SU(3)_H$ with the LH fermions of the three families transforming as triplets 
 and the RH ones as anti-triplets, as it is motivated by the grand unification.  

However, in the Standard Model framework  one has more possibilities. 
Namely, in the limit of  vanishing Yukawa couplings $Y_{u,d,e} \to 0$ in (\ref{Yukawa-SM}),   
the SM Lagrangian acquires a maximal global chiral symmetry 
$U(3)_Q \times U(3)_u \times U(3)_d  \times U(3)_\ell \times U(3)_e$   
under which fermion species $Q$, $u$ etc. transform  
as triplets of independent $U(3)$ groups.  
It is tempting to consider that the non-abelian $SU(3)$ factors of this maximal flavor  symmetry 
are related to  gauge symmetries.\footnote{Gauging of chiral $U(1)$ factors is problematic 
because of anomalies. In fact,  one combination of $U(1)$ factors can be rendered practicable 
via the Green--Schwarz mechanism and 
there are fermion mass models in which such 
anomalous  gauge symmetry $U(1)_A$ is used as a flavor symmetry  \cite{U1A}. }  


Let us concentrate on the lepton sector and discuss the gauge symmetry 
$SU(3)_\ell \times SU(3)_e$ \cite{Benedetta}
under which the LH and RH lepton fields transform as 
\be{reps} 
\ell_{L\alpha} = \dub{\nu_\alpha}{e_\alpha}_L\!\! \sim (\mathbf{3}_\ell,1),   
\quad  e_{R\gamma} \sim (1,\mathbf{3}_e)  
\ee 
where 
$\alpha=1,2,3$ and $\gamma=1,2,3$ are the indices of $SU(3)_\ell$ and $SU(3)_e$ respectively.  
This set of fermions is not anomaly free. The ways of the anomaly cancellation 
were discussed in Ref. \cite{Benedetta} and in this 
letter we shall not concentrate on this issue. 


For breaking $SU(3)_\ell\times SU(3)_e$ we introduce flavon fields,  three triplets 
$\eta_{i\alpha}$ of $SU(3)_\ell$ and three triplets $\xi_{i\gamma}$ of $SU(3)_e$, $i=1,2,3$.  
Then the  charged lepton masses emerge from the gauge invariant dimension--6  
operator 
\be{op-leptons} 
\frac{y_{ij} }{\cM^2}\,  \eta_{i\alpha} \ov\xi_{j}^{\gamma} \phi \, \overline{ \ell_{L\alpha}  } 
 e_{R\gamma} \, + \, {\rm h.c.}
\ee 
where $y_{ij}$ are order one constants, $\phi$ is the Higgs doublet  and  $\cM$ is a cutoff scale. 
In an UV-complete theory such operators can be induced via seesaw-like mechanism 
by integrating out some heavy scalar or fermion states \cite{Chkareuli,PLB83}. 
However,  concrete model building is not the scope of this paper, and for our demonstration   
effective operator analysis is sufficient.   As for the neutrinos,  their Majorana masses 
are  induced by the higher order operator
\be{op-nu} 
\frac{h_{ij}  }{\cM_\nu^3} \,\ov{\eta}_{i}^{\alpha}\ov{\eta}_{j}^{\beta} \,  \phi \phi   
\, \ell^T_{L\alpha} C \ell_{\beta}   \, + \, {\rm h.c.}  \, 
\ee
where $h_{ij} = h_{ji}$. 
The cutoff scale $\cM_\nu$ of this operator is not necessarily the same as the scale $\cM$ of 
operator (\ref{op-leptons}).

In order to generate non-zero masses of all three leptons $e,\mu,\tau$,  
all three $SU(3)_\ell$ flavons $\eta_i$ as well as $SU(3)_e$  $\xi_i$ should have non-zero VEVs  
with disoriented directions.  
This means that the VEVs $\langle \eta_{i\alpha} \rangle$ 
should form a rank-3 matrix. 
Without losing generality, the flavon basis can be chosen 
so that the matrix $\langle \eta_{i\alpha} \rangle$ is diagonal, 
$\langle \eta_{i\alpha} \rangle = w_i \delta_{i\alpha}$, 
i.e. the flavon VEVs are orthogonal:   
\be{VEV}
\langle \eta_1 \rangle = \left(\begin{array}{c}
w_1 \\ 0 \\ 0 
\end{array}\right), ~
\langle \eta_2 \rangle = \left(\begin{array}{c}
0 \\ w_2 \\ 0
\end{array}\right)  , ~
\langle \eta_3 \rangle = \left(\begin{array}{c}
0 \\ 0 \\ w_3
\end{array}\right)  
\ee 
%
Analogously, for $\xi$--flavons we take $\langle \xi_{i\gamma} \rangle = v_i \delta_{i\gamma}$. 
After plugging these VEVs into (\ref{op-leptons}) we obtain the  leptonic Yukawa matrices in the SM Lagrangian 
(\ref{Yukawa-SM}) as
\be{Ye} 
Y_e^{ij} = y_{ij} \frac{w_i v_j}{\cM^2} 
\ee
Since the couplings (\ref{op-leptons})  should give the lepton mass hierarchy, 
we consider that the latter emerges due to the VEV hierarchy 
$v_3 \gg v_2 \gg v_1$ in $SU(3)_e$ symmetry breaking, 
i.e.  $v_3 : v_2 : v_1 \sim m_\tau : m_\mu : m_e$  as it is described in Ref. \cite{Benedetta}.
On the other hand, operator (\ref{op-nu}) should give the observed neutrino mass pattern, 
$m_\nu^{ij} = h_{ij} w_i w_j v_{\rm w}^2/\cM_\nu^3$, 
and in particular the large neutrino mixing. This implies that   
 $SU(3)_\ell$ breaking flavons $\eta$ 
should have comparable VEVs,  $w_3\sim w_2 \sim w_1$.

Gauge bosons $\cF^\mu_a$ of $SU(3)_\ell$,  associated to the Gell-Mann matrices 
$\lambda_a$, $a=1,2,...8$,  interact as $g \cF^\mu_a J_{a\mu}$ 
with the respective currents $J_{a\mu} = J_{a\mu}^{(e)} + J_{a\mu}^{(\nu)}  
= \frac12 \ov{\boldsymbol{e}_L} \gamma_\mu  \lambda_a {\boldsymbol{e}_L} 
+  \frac12  \ov{\boldsymbol{\nu}_L} \gamma_\mu  \lambda_a  {\boldsymbol{\nu}_L}  $,
where $g$ is the gauge coupling constant, 
${\boldsymbol{e}_L} = (e_1,e_2,e_3)_L^T$ 
and  ${\boldsymbol{e}_L} = (\nu_1,\nu_2,\nu_3)_L^T$ respectively
denote the family triplets of the LH charged leptons and neutrinos.

At low energies these couplings induce four-fermion (current $\times$ current)  interactions: 
\be{CC}
\mathcal{L}_{\rm eff} = - \frac{g^2}{2} J_{a}^{\mu}\, \left(M^2\right)^{-1}_{ab} \, J_{b\mu}  
\ee
where $M^2_{ab}$ is the squaredmass matrix of gauge bosons $\cF^\mu_a$ which 
in the flavon VEV basis (\ref{VEV})  is essentially diagonal 
apart of  a non-diagonal $2\times 2$ block related to ${\cal F}^\mu_3$ - ${\cal F}^\mu_8$ mixing.
Namely, the masses of $\cF^\mu_{1,2}$, $\cF^\mu_{4,5}$ and  $\cF^\mu_{6,7}$ 
are 
\beqn{67}
&& M_{1,2}^2 = \frac{g^2}{2}(w_2^2 + w_1^2) =  \frac{g^2}{2}v_\cF^2 ,     \\
&& M_{4,5}^2 = \frac{g^2}{2}(w_3^2 + w_1^2),  \quad M_{6,7}^2 = \frac{g^2}{2}(w_3^2 + w_2^2)\, . 
\nonumber 
\eeqn
 As for  ${\cal F}^\mu_3$  and ${\cal F}^\mu_8$ they have a mass mixing and 
 their mass matrix reads 
\be{38} 
M_{38}^2  =  \frac{g^2}{2} 
\mat{ w_2^2 + w_1^2} {\frac{1}{\sqrt3}(w_1^2 - w_2^2)}
 {\frac{1}{\sqrt3}(w_1^2 - w_2^2)} {\frac13(4 w_3^2 + w_1^2 + w_2^2) } \, .
 \ee
Notice that if $w_1=w_2=v_\cF/\sqrt2$, this matrix becomes diagonal.
 In the following, for the simplicity of our demonstration, 
we analyze this case.\footnote{
Similar analysis can be done also for a general case $w_1\neq w_2$,  
along the lines of Ref. \cite{Benedetta} where such analysis was done for 
the RH gauge sector $SU(3)_e$.   
} 
Then for the gauge boson masses  we have 
$M_a^2 = (g^2/2) (x_a v_\cF)^2$, where 
\be{x}
x^2_{1,2,3}  = 1, \quad  x^2_{4,5,6,7}  = \frac{r+1}{2}  ,  \quad x^2_8 = \frac{2r+1}{3} 
\ee
and $r = 2 w_3^2/v_\cF^2$.  
Then operators (\ref{CC})  
can be rewritten as $\mathcal{L}_{\rm eff} = \mathcal{L}_{\rm eff}^{e\nu} + 
\mathcal{L}_{\rm eff}^{ee} + \mathcal{L}_{\rm eff}^{\nu\nu}$ where 
 \begin{eqnarray} \label{CXC}  
&& \mathcal{L}_{\rm eff}^{e\nu}  = - \frac{2G_\cF}{\sqrt2 }\, \sum_{a=1}^8 
\big(\ov{\boldsymbol{e}_L} \, \gamma^\mu  \frac{\lambda_a}{x_a} \, {\boldsymbol{e}_L}\big) 
\big(\ov{\boldsymbol{\nu}_L} \, \gamma_\mu \frac{\lambda_a}{x_a} \,  {\boldsymbol{\nu}_L} \big) 
\nonumber   \\ 
&&  \mathcal{L}_{\rm eff}^{ee} =  - \frac{G_\cF}{\sqrt2}\, \sum_{a=1}^8  
\big(\ov{\boldsymbol{e}_L} \, \gamma_\mu \frac{\lambda_a}{x_a} \, {\boldsymbol{e}_L} \big)^2 \\
&& \mathcal{L}_{\rm eff}^{\nu\nu} =  - \frac{G_\cF}{\sqrt2}\, \sum_{a=1}^8  
\big(\ov{\boldsymbol{\nu}_L}\, \gamma_\mu \frac{\lambda_a}{x_a} \, {\boldsymbol{\nu}_L} \big)^2 
\nonumber 
\end{eqnarray} 
where  $4G_\cF/\sqrt2 = 1/v_\cF^2$. 
Obviously, the factor $g^2/2$ in operators cancels out and the strength 
of these operators is determined solely by the VEVs (\ref{VEV}). 

The first term $\mathcal{L}_{\rm eff}^{e\nu}$ contains operator (\ref{new}) which 
contributes to the muon decay $\mu \to e \nu_\mu \bar{\nu}_e$ as $G_\mu = G_F + G_\cF$. 
It  is induced   by exchange of gauge bosons $\cF_1^\mu$ and $\cF_2^\mu$,  
or more precisely by the combination $(\cF_1^\mu \pm i\cF_2^\mu)/\sqrt2$, 
as in second diagram of Fig.~\ref{diagrams}. 
As it was pointed out in previous section, 
for restoring the CKM unitarity one needs $\delta_\mu = G_\cF/G_F = (v_{\rm w}/v_\cF)^2$ 
to be around $ 7\times 10^{-4}$ which corresponds 
to the flavor scale $v_\cF \approx 6\div 7$~TeV. 

The similar operators in $\mathcal{L}_{\rm eff}^{e\nu}$ mediated by 
the gauge bosons $\cF_{4,5}^\mu$ and $\cF_{6,7}^\mu$ 
contribute to the taon leptonic decays 
$\tau \to e \nu_\tau \bar{\nu}_e$ and $\tau \to \mu \nu_\tau \bar{\nu}_\mu$. 
Then, in the case $w_{1,2,3}\sim v_\cF$  but $w_1\neq w_2$,   the branching ratio 
$\Gamma(\tau \to \mu \nu_\tau \bar{\nu}_\mu)/\Gamma(\tau \to e \nu_\tau \bar{\nu}_e)$ 
can have up to $O(10^{-3})$ deviation from the SM prediction $0.9726$ which can be 
experimentally testable.  
(For a comparison, the present experimental value of this ratio is 
$0.9762(28)$ \cite{PDG2018}.)  
In addition, in $\mathcal{L}_{\rm eff}^{e\nu}$ the terms with the ``diagonal' 
generators $\lambda_3$ and $\lambda_8$ give rise also the 
non-standard neutrino interactions with leptons with coupling constants 
$\sim G_\cF = \delta_\mu G_F$, well below the experimental constraints.  

The last term $\mathcal{L}_{\rm eff}^{\nu\nu}$ in (\ref{CXC}) contains the 
non-standard interactions between neutrinos, but present experimental limits on 
such interactions are rather weak. 
On the other hand, the second term $\mathcal{L}_{\rm eff}^{ee}$ in (\ref{CXC}) 
containing charged leptons in principle is testable for the scale $v_\cF$ of several TeV.  

Interestingly, if  the flavor eigenstates $e_{1},e_{2},e_{3}$ 
are the mass eigenstates $e,\mu,\tau$, the terms (\ref{CXC}) do not contain 
any LFV operators inducing processes like $\mu \to 3e$, $\tau \to 3\mu$ etc. 
However, the lepton flavor-conserving contact operators 
$-\frac{4\pi}{\Lambda_L^2}(\ov{e_L}\gamma_\mu e_L)^2$, 
$-\frac{2\pi}{\Lambda_L^2}(\ov{e_L}\gamma^\mu e_L)(\ov{\mu_L}\gamma_\mu \mu_L)$, etc. 
are restricted by the `compositeness' limits 
$\Lambda^-_L(eeee) > 10.3$~TeV and $\Lambda^-_L(ee\mu\mu) > 9.5$ TeV.
Comparing these operators with the corresponding terms in (\ref{CXC}) 
and taking into account the relations (\ref{x}),
the `compositeness' scales can be expressed in terms of the scale $v_\cF$.    
Hence, we obtain the limit 
\be{compo} 
v_\cF > \left(\frac{r+1}{r+0.5} \right)^{1/2} \times 2.1 ~ {\rm TeV} \, . 
\ee
Here the $r$--dependent pre-factor approaches 1 when $r\gg 1$ 
and it becomes $\sqrt2$ in the opposite limit $r \ll 1$.  Thus, the strongest 
limit  emerges in the latter case, $v_\cF > 3$~TeV or so, which is anyway   
fulfilled for our benchmark range $v_\cF \simeq (6\div7)$~TeV. 

The flavor eigenstates $e_{1},e_{2},e_{3}$ 
coincide with the mass eigenstates $e,\mu,\tau$, if the Yukawa matrix $Y_e^{ij}$ 
in (\ref{Ye}) is diagonal. This can be achieved by imposing some additional discrete 
symmetries between the flavons $\eta_i$ and $\xi_i$ of $SU(3)_\ell$ and $SU(3)_e$ sectors 
which would forbid the non-diagonal terms $y_{ij}$ in operator (\ref{op-leptons}).  
However, in general case the initial flavor basis of the LH leptons is related to  
the mass basis by the unitary transformation
\begin{equation}\label{VL}
\left(\begin{array}{c}
e_1\\e_2\\e_3
\end{array}\right)_{\!\!\!L} 
\!\! = U_L \! \left(\begin{array}{c}
e\\ \mu \\ \tau \end{array}\right)_{\!\!\!L} \! 
=\left(\begin{array}{ccc}
U_{1e} & U_{1\mu} & U_{1\tau} \\ U_{2e} &U_{2\mu}& U_{2\tau}\\ U_{3e}& U_{3\mu}& U_{3\tau}
\end{array}\right) \!\!
\left(\begin{array}{c}
e\\ \mu \\ \tau
\end{array}\right)_{\!\!\!L}
\end{equation}
Then, in the basis of mass eigenstates, the operators $\mathcal{L}_{\rm eff}^{ee}$ 
read as in (\ref{CXC}) but with the substitution $\lambda_a/x_a \to U^\dagger (\lambda_a/x_a) U$.  
Interestingly, in the limit $r=1$, i.e. when the VEVs $w_{1,2,3}$ are equal  
and so $x_a =1$, all flavor bosons $\cF_a^\mu$ have equal masses,  
and the substitution $\lambda_a \to U^\dagger \lambda_a U$ is simply a 
basis redetermination of the Gell-Mann matrices. Therefore, no LFV effects will emerge 
in this case since the global $SO(8)_\ell$ symmetry acts as a custodial symmetry. 
Namely, by Fierz transformations, using also the Fierz identities for the Gell-Mann matrices, 
we obtain
\be{fierz} 
- \frac{G_\cF}{\sqrt2}\, \sum_{a=1}^8 
\big(\ov{\boldsymbol{e}_L} \gamma_\mu \lambda_a  {\boldsymbol{e}_L} \big)^2 = 
- \frac{4}{3}  \frac{G_\cF}{\sqrt2}\, \big(\ov{\boldsymbol{e}_L} \gamma_\mu {\boldsymbol{e}_L} \big)^2 
\ee
Obviously, the latter expression is invariant under the unitary transformation (\ref{VL}). 

In general case $r\neq 1$, the mixing (\ref{VL}) gives rise to the LFV operators as e.g. 
the one inducing $\mu\to 3e$ decay:
\beqn{mu3e} 
&& -\frac{4G_{\mu eee} }{\sqrt2} \big(\ov{e_L} \gamma^\mu \mu_L\big)\big(\ov{e_L} \gamma^\mu e_L\big)  
\, + \, 
{\rm h.c.} \, ,    \nonumber \\
&& \frac{4G_{\mu eee} }{\sqrt2} = 
\frac{C(r) }{2v_\cF^2} \left[1 + \frac{1-r}{r} \vert U_{3e} \vert^2 \right] U^\ast_{3e} U_{3\mu} \, , 
\eeqn
where the function $C(r) = (r-1) r \big[(r+1)(r+0.5)\big]^{-1}$ is limited as $\vert C(r) \vert < 1$, 
reaching the maximal value at  $r\gg 1$, and it vanishes at $r=1$. 
Then, taking $\vert U_{3e} \vert \ll 1$, we obtain for the branching ratio of $\mu \to 3e$ decay 
\be{branching}
\frac{\Gamma(\mu \to ee\bar{e})}{\Gamma(\mu \to e\nu_\mu\bar{\nu}_e)} = 
\frac12 \left\vert \frac{G_{\mu eee}}{G_F} \right\vert^2 = \frac{1}{8} \left(\delta_\mu 
C(r) \vert U^\ast_{3e} U_{3\mu}  \vert \right)^2 
\ee
The experimental upper bound on this branching ratio is $10^{-12}$ \cite{PDG2018}.  
Taking $\delta_\mu = (v_{\rm w}/v_\cF)^2 = 7\times 10^{-4}$, the limit 
$\delta_\mu \vert C U^\ast_{3e} U_{3\mu}  \vert/\sqrt8 < 10^{-6}$ translates 
into $\vert C U^\ast_{3e} U_{3\mu}  \vert < 0.4 \times 10^{-2}$ which is  nicely 
satisfied if the lepton mixing angles in (\ref{VL}) are comparable with the CKM mixing angles 
in (\ref{CKM}) or even larger.  E.g. if the VEV ratio is in between $r=0.5\div 1.5$, 
then $\vert C(r) \vert < 1/7$ so that $\vert U^\ast_{3e} U_{3\mu}  \vert < (1/6)^2$ or so 
would suffice for properly suppressing the $\mu\to 3e$ decay rate. This means that 
in this case the matrix elements $\vert U_{3\mu}  \vert$ and $\vert U_{3e} \vert $ 
can be almost as large as the Cabibbo angle $\sin\theta_C = V_{us}$.  
The experimental limits on other LFV effects as e.g. $\tau\to 3\mu$ are weaker, 
and following the lines of Ref. \cite{Benedetta} one can show that in our model 
with 
$v_\cF \simeq 6$ TeV or so, 
they are fulfilled even for whatever large mixings  in (\ref{VL}). 
Once again, for $r=1$ all LFV effects are vanishing  
owing to custodial symmetry, see Eq. (\ref{fierz}).


\medskip
\noindent
{\bf 7.} Let us discuss briefly how the hypothesis $G_\mu \neq G_F$ could 
affect the SM precision tests. In the SM, at tree level, the weak gauge boson masses 
are $M_W= gv_{\rm w}/\sqrt2=e v_{\rm w}/\sqrt2 \sin\theta_W$ 
and $M_Z=M_W/\cos\theta_W$ where $\theta_W$ is the weak angle. 
For precision tests the radiative corrections are important which depend also 
on the top quark and Higgs mass. 

The world averages of experimentally measured masses of $Z$ and $W$ 
reported by PDG 2018 are \cite{PDG2018}: 
\beqn{MWZ-exp} 
&& M_Z^{\rm exp}  = 91.1876(21)\,{\rm GeV},   \nonumber \\
&&  M_W^{\rm exp} = 80.379(12)\,{\rm GeV}, 
\eeqn
while the SM global fit yields to the following values:
\beqn{MWZ-SM} 
&& M_Z^{\rm SM} = 91.1884(20)~{\rm GeV},  \nonumber \\
&&    M_W^{\rm SM} = 80.358(4)~{\rm GeV}. 
\eeqn
Hence, the theoretical and experimental values of $Z$-mass are in perfect agreement 
while for $W$-boson the two values have about  $1.6\sigma$ discrepancy: 
\be{DeltaMW} 
M_W^{\rm exp} - M_W^{\rm SM}= (21 \pm 13)~{\rm MeV} 
\ee 

In the SM the mass of $W$-boson, including radiative corrections, is determined as 
\be{MW} 
M_W = \frac{A_0}{\hat{s}_Z  (1-\Delta\hat r_W)^{1/2} } 
\ee 
where $A_0 = (\pi\alpha/\sqrt2 G_F)^{1/2} = 37.28039(1)$~GeV taking  
$G_F=G_\mu$,  
the factor $1-\Delta\hat r_W=0.93084(8)$ includes the main radiative corrections   
and $\hat{s}_Z^2 =1.0348(2) s_W^2$ is the corrected value of 
$\sin^2 \theta_W(M_Z)$  by including the top and Higgs mass dependent corrections.
The theoretical mass $M_W=80.358(4)$~GeV (\ref{MWZ-SM}) is then obtained 
by substituting  in (\ref{MW}) the value $\hat{s}_Z^2 = 0.23122(3)$ obtained from the SM global fit 
 \cite{PDG2018}. 
In our scenario, however, $G_F \neq G_\mu$. 
Should we just set in $A_0$ instead of $G_F=G_\mu$ the ``corrected" value 
$G_F = (1+ \delta_\mu)^{-1} G_\mu$,  
then $A_0$ should be rescaled by a factor $(1+\delta_\mu)^{1/2}$, and correspondingly 
the ``theoretical" value of $M_W$ (\ref{MW}) too. 
In particular, for $\delta_\mu = 7\times 10^{-4}$ we would get 
$M_W=80.386$~GeV, right in the ball-park  of the experimental values (\ref{MWZ-SM}). 
However, this is not the right thing to do. 

In the global fit of SM $M_Z$ is one of the input parameters with smallest experimental errors, 
along with the fine structure constant $\alpha$ and the ``muon" Fermi constant $G_\mu$.  
Essentially, this is the main reason of the good coincidence between $M_Z^{\rm exp}$ and 
$M_Z^{\rm SM}$.  In fact, the SM implies the relation
\be{MZ} 
M_Z = \frac{M_W}{ \hat{c}_Z \hat{\rho}^{1/2}} = 
\frac{A_0}{\hat{s}_Z \hat{c}_Z  (1-\Delta\hat r_W)^{1/2} \hat{\rho}^{1/2} } 
\ee 
where $\hat{\rho}=1+\rho_t + \delta\rho= 1.01013(5)$ includes the weak isospin breaking effects, 
dominantly from the quadratic $m_t$ dependent corrections $\rho_t = 3G_Fm_t^2/8\sqrt2\pi^2$. 
Therefore, taking the experimental value of $Z$-mass (\ref{MWZ-exp}), 
Eq. (\ref{MZ}) can be used for determination of $\hat{s}_Z^2$ parameter,  
$\hat{s}_Z^2=0.23123(3)$.  This, in turn, from $M_W= M_Z \hat{\rho}^{1/2} \hat{c}_Z$
 gives $M_W=80.357(4)_{\rm SM}$~GeV, i.e. practically the same as the global fit result (\ref{MWZ-SM}).  
This is because the determination of the parameter $\hat{s}_Z^2$ in the SM global fit 
is dominated by the results of $Z$-pole measurements. 

However, in our scenario rescaling $A_0 \to A_0(1+\delta_\mu)^{1/2}$ 
changes the value of  $\hat{s}_Z^2$. 
In particular, taking $\delta_\mu = (7.6\pm1.6)\times 10^{-4}$, we get 
$\hat{s}_Z^2=0.23148(3)_{\rm SM}(5)_{\delta_\mu}$. Then, again from 
$M_W= M_Z \hat{\rho}^{1/2} \hat{c}_Z$, we get $M_W=80.344(4)_{\rm SM}(3)_{\rm \delta_\mu}$~GeV. 
Thus, unfortunately, while the effect is there, in reality it goes right to the opposite direction. 
So, our determination of $M_W$ differs from $M_W^{\rm SM}$, 
$M_W^{\rm SM} - M_W^{\rm our} = (13 \pm 3)$~MeV. 
Thus, with $M_W^{\rm SM}$ already being in tension with the experimental value 
(\ref{MWZ-exp}), our result has more tension:    
$M_W^{\rm exp} - M_W^{\rm our}= (35 \pm 13)$~MeV ($2.7\sigma$). 
However, let us remark that the tension with the 
latest  results of ATLAS $M_W^{\rm ATL} = 80.370(19)$ is less, 
$M_W^{\rm ATL} - M_W^{\rm our}= (26 \pm 20)$~MeV ($1.3\sigma$). 
If the tension will increase with future precision, this would mean that one has to admit 
at least some minimal step beyond the SM.  The relation between $W$ and $Z$ masses 
can be improved by increasing of $\rho$-parameter via e.g. the VEV  $\sim 1$ GeV of a
scalar triplet of the electroweak $SU(2) \times U(1)$, or by diminishing $Z$ mass by few MeV
e.g. via  its mixing with some extra gauge bosons like $Z'$ or perhaps also with 
the flavor gauge bosons considered in the previous section. 


\medskip 

\noindent {\bf 8.} The value  $\vert V_{ud} \vert$ can be extracted also from free neutron decay 
by combining the results on the measurements of the neutron lifetime $\tau_n$ 
with  those of the axial current coupling constant $g_A=G_A/G_V$. The master formula reads 
(see e.g. in a recent review \cite{Gonzalez-Alonso:2018}): 
\beqn{Vud-n}
&& \vert V_{ud} \vert^2 = \frac{K/ \ln 2 }{ G_F^2 \cF_n \tau_n \, (1+3g_A^2)  (1+ \Delta_R^V)} \nonumber \\
&& \quad \quad ~~ =  \frac{5024.46(30)~{\rm s}}{\tau_n (1+3g_A^2)(1+ \Delta_R^V)} 
\eeqn
where $\cF_n = f_n(1+\delta'_R)$ is the neutron $f$-value $f_n= 1.6887(1)$ 
corrected by the long-distance QED correction $\delta'_R = 0.01402(2)$ \cite{Hardy3}.  
This equation, taking the values $\tau_n=880.2\pm 1.0$ s and $g_A=1.2724\pm 0.0023$ 
adopted in PDG 2018 \cite{PDG2018}, would give the value 
$\vert V_{ud} \vert = 0.97577(55)_{\tau_n} (146)_{g_A}(18)_{\Delta_R^V}  = 0.97577(157) $.
It is compatible with $ \vert V_{ud} \vert = 0.97370(10)_{\cF t} (10)_{\Delta_R^V} = 0.97370(14)$
obtained from (\ref{Vud-super}) and used in (\ref{data-PDG}),  
but has an order of magnitude larger error. 

\begin{figure}
\includegraphics[width=0.45\textwidth]{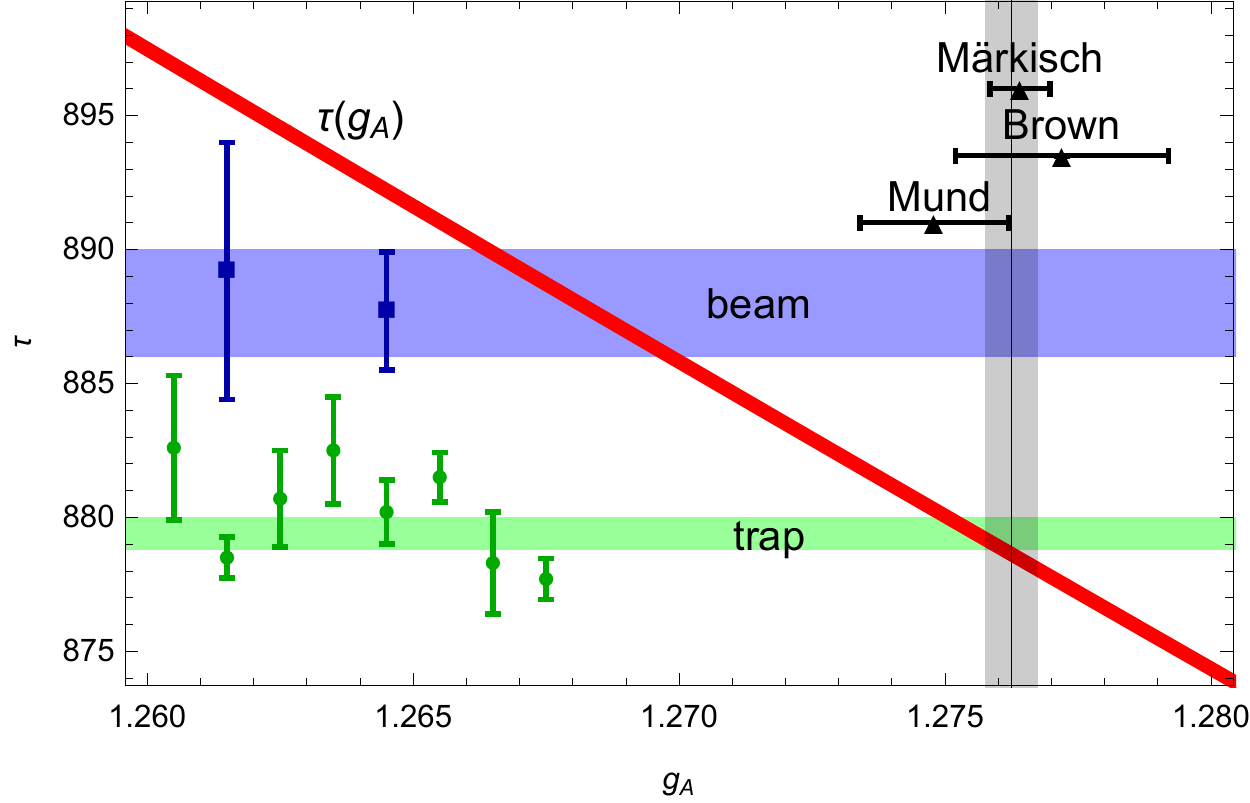}
\caption{The red band shows the precision relation (\ref{tau-g} between $g_A$ and $\tau_n$.  
Black triangles with horizontal error bars show values of $g_A$ 
reported in Refs. \cite{Mund,UCNA,Markisch} and vertical grey band corresponds to their 
average (\ref{gA}).
Green circles  show values of $\tau_n$ reported by trap experiments  
\cite{Mampe:1993,Serebrov:2005,Pichlmaier:2010,Steyerl:2012, Arzumanov:2015,Serebrov:2017,Ezhov:2014,Pattie:2017} with respective error bars
and horizontal green band shows their average (\ref{trap}).  
Blue squares and blue horizontal band show the the same for beam 
experiments \cite{Byrne:1996,Yue:2013}. 
 }
\label{fig:n-life}
\end{figure}

However, rather than for determination of $\vert V_{ud} \vert$,  Eq. (\ref{Vud-n}) 
can be used for  a consistency check. 
Namely, by comparing it with Eq. (\ref{Vud-super}) 
 we get a relation between $\tau_n$ and $g_A$ \cite{Czarnecki}:
\be{tau-g} 
\tau_n = \frac{2 \cF t}{\ln 2 \, \cF_n (1+3g_A^2)} =  \frac{5172.0(1.1) ~{\rm s} }{1+3g_A^2} 
\ee
%
In Fig.~\ref{fig:n-life} this relation  is shown by the red band. 
This formula is very accurate since the common factors in Eqs.  (\ref{Vud-super})  and 
(\ref{Vud-n}) cancel out,  including the Fermi constant and radiative corrections  $\Delta_R^V$.  

For the axial current coupling $g_A$, the PDG 2018 quotes a value $g_A=1.2724\pm 0.0023$. 
However, the results of the latest and most recent experiments  
\cite{Mund,UCNA,Markisch} which measured $\beta$-asymmetry parameter
using different techniques    (the cold neutrons in PERKEO II and PERKEO III 
experiments \cite{Mund,Markisch}  and 
ultra-cold neutrons in the UCNA experiment \cite{UCNA}),   
are in perfect agreement among each other, 
and their average determines the axial current coupling  $g_A$ with 
impressive (better than one per mille)  precision: 
\be{gA}
g_A = 1.27625 \pm 0.00050\, .
\ee 
Fig.~\ref{fig:n-life} shows the results of Refs. \cite{Mund,UCNA,Markisch} and their 
average (vertical grey band).  For $g_A$ in this range  Eq. (\ref{tau-g}) gives 
the Standard Model prediction for the neutron lifetime  
\be{tauSM}
\tau_n^{\rm SM} = 878.7 \pm 0.6~{\rm s}  
\ee

From the experimental side, the neutron lifetime is measured in two types of experiments. 
The trap experiments measure the disappearance rate of the ultra-cold neutrons (UCN)  
by counting the survived neutrons after storing them for different times  in the UCN traps 
 and  determine the neutron decay width $\Gamma_n = \tau_n^{-1}$.  
The beam experiments are the appearance experiments, 
measuring the width of $\beta$-decay $n\to pe\bar\nu_e$, $\Gamma_\beta=\tau_\beta^{-1}$,  
by counting the produced protons  in the monitored beam of cold neutrons. 
In the Standard Model the neutron decay should always  produce  a proton,  
and so both methods should measure the same value  $\Gamma_n = \Gamma_\beta$.

However,  there is  tension between the results obtained using two different methods, 
as it  was pointed out in Refs. \cite{Serebrov:2011}. 
Fig.~\ref{fig:n-life} clearly demonstrates the discrepancy. 
Namely, by averaging the presently available results of eight  trap experiments  
\cite{Mampe:1993,Serebrov:2005,Pichlmaier:2010,Steyerl:2012, Arzumanov:2015,Serebrov:2017,Ezhov:2014,Pattie:2017}
one obtains:
 \be{trap} 
\tau_{\rm trap} = 879.4 \pm 0.6~{\rm s}  \, , 
\ee 
which is compatible with the SM prediction (\ref{tauSM}).
On the other hand,  the beam experiments  \cite{Byrne:1996,Yue:2013}   
yield   
\be{beam}
\tau_{\rm beam} = 888.0 \pm 2.0~{\rm s} \, .
\ee
which is  about $4.4\sigma$ away from the SM predicted 
value (\ref{tauSM}).\footnote{The PDG 2018 average $\tau_n = 880.2 \pm 1.0$~s  includes  
the results of two beam experiments \cite{Byrne:1996,Yue:2013} and five trap experiments 
\cite{Mampe:1993,Serebrov:2005,Pichlmaier:2010,Steyerl:2012, Arzumanov:2015},  
with the error rescaled up by a factor 
$\sqrt{\chi^2_{\rm dof} } \approx 2$ for a loose compatibility between the data, essentially 
between the trap and beam experiments. Results of three recent trap experiments 
\cite{Serebrov:2017,Ezhov:2014,Pattie:2017} published in 2018 were not included. } 

Therefore, due to consistency with the SM prediction (\ref{tau-g}), 
it is more likely that the true value of the neutron lifetime is the one 
measured by trap experiments (\ref{trap}). 
About 1  per cent deficit of produced protons in the beam experiments \cite{Byrne:1996,Yue:2013} 
might be due to some unfixed systematic errors. 
Alternatively, barring the possibility of uncontrolled systematics and considering the problem as real, 
a new physics  must be invoked which could explain about one per cent deficit 
of protons produced in the beam experiments. One interesting possibility can be 
related to the neutron--mirror neutron ($n-n'$) oscillation \cite{BB-nn'}, provided that ordinary and mirror 
neutrons have a tiny mass difference 100 neV or so \cite{Berezhiani-nn}. 
Then in large magnetic fields (5 Tesla or so) used in beam experiments $n-n'$ conversion 
probability can be resonantly enhanced to about $\sim 0.01$ and thus corresponding 
fraction of neutrons converted in mirror neutrons will decay in an invisible (mirror) channel 
without producing ordinary protons. 

Concluding this section, let us remark that the present precision calculation 
of the short-range radiative corrections $\Delta_R^V$ \cite{Seng:2018} and respective 
redetermination of $V_{ud}$ has no influence on the determination of the neutron 
lifetime (\ref{tauSM}) obtained from Eq. (\ref{tau-g}) 
which in fact directly relates the value of $\tau_n$ 
to the value $\cF t$ accurately measured in superallowed $0^+-0^+$ nuclear transitions 
and to the value $g_A=G_A/G_V$ obtained from accurate measurements of $\beta$-asymmetry. 
Notice that the relation (\ref{tau-g}) remains valid also in the presence of non-standard 
vector or axial interactions contributing to the neutron decay, since the value of $G_V$ 
(independently whether it is equal to $G_F \vert V_{ud} \vert$ or not) 
anyway cancels out \cite{LHEP} 
and only the ratio $g_A=G_A/G_V$ remains relevant which value is accurately 
determined from the measurements of $\beta$-asymmetry.  
In particular, Eq. (\ref{tau-g}) remains valid in our model with  
$G_F\neq G_\mu$ discussed  in previous section, or more generically 
for any modification of the SM introducing new vector and axial couplings 
contributing in operator (\ref{ud}). 

\begin{table*}
\begin{tabular}{c c  c c c }
\hline
\hline
\rule[-2mm]{0mm}{6mm}
  & ~ CKM [PDG] ~ & ~ CKM [post 2018] ~ &  ~ CKM$+ \, b'$ ~ &  ~ CKM$+ \cF$   \\
\hline
\rule[-2mm]{0mm}{5mm}
C &  $0.2257(9) $  & ~ $0.22780(60)$ ~ & ~$0.22443(61)$~ & ~ $0.22460(61)$ ~\\
\rule[-2mm]{0mm}{4mm}
B & $0.2256(10)$  & $0.22535(45)$ & $0.22518(45)$ & $0.22535(45)$ \\
\rule[-2mm]{0mm}{4mm}
A & $0.2238(8)$ & $0.22333(60)$ & $0.22333(60)$ & $0.22350(60)$  \\
\rule[-2mm]{0mm}{4mm}
$\overline{A\!+\!B}$ & $0.2245(6)$ & $0.22463(36)$ & $0.22452(36)$ & $0.22469(36)$ \\
\hline
\rule[-2.mm]{0mm}{6mm}
$\overline{A\!+\!B\!+\!C}$ ~ & $0.2248(5)$  & $0.22546(31)$ & $0.22449(31)$ &  $0.22467(31)$  \\
& $\chi^2=3.4$ & $\chi^2=27.7$ $^\dagger$ & $\chi^2=6.1$ & $\chi^2=6.1$ \\
\hline
\rule[-2.mm]{0mm}{6mm}
$\vert V_{us} \vert $  ~ & $0.2248(7)$   & $0.2255(12)$ $^\dagger$  & $0.2245(5)$ &  $0.2247(5)$  \\
\rule[-2.mm]{0mm}{4mm}
$\vert V_{ud} \vert $  ~ & $0.97440(16)$   & $0.97424(27)$ $^\dagger$  & $0.97369(12)$ &  $0.97443(12) $  \\
\hline
\hline
\end{tabular}
\caption{\label{Table} The 1st column shows independent $\vert V_{us} \vert$ determinations A, B, C 
from the PDG dataset (\ref{data-PDG}) by assuming 3-family CKM unitarity (\ref{unitarity}), 
their averages and total $\chi^2$ value. The last two rows show the conservative estimation of 
$\vert V_{us} \vert $ with error-bar rescaled by $\sqrt{\chi^2_{\rm dof}}$ and the corresponding value of 
$\vert V_{ud} \vert $.   
Other columns  show the same but obtained from 
after 2018 dataset (\ref{data-new}) by assuming respectively  3-family CKM unitarity (\ref{unitarity}), 
unitarity extended to 4th quark $b'$ with $\vert V_{ub'} \vert =0.04$,  
and 3-family CKM but taking $G_\mu/G_F=1+\delta_\mu$ with $\delta_\mu=7.6\times 10^{-4}$. 
Mark $^\dagger$  in 2nd column indicates that for that large $\chi^2$ 
the error-rescaling  by $\sqrt{\chi^2_{\rm dof}}=3.7$ does not make much sense since the data 
are incompatible.  
  }
\end{table*}


\medskip
\noindent
{\bf 9.}  As concluding remarks. 
The present experimental and theoretical 
accuracy in independent determination of the first row elements of the CKM matrix 
indicates towards about 4.4$\sigma$ deviation from the unitarity (\ref{CKM}). 
This can be indication to the new physics at the scale of few TeV. 
We investigated two possible scenarios in order to fill the gap. 
The respective results are summarised it Table \ref{Table}. 

The first, rather straightforward possibility is related to the existence of ``fourth family" 
in the form of a vector-like couple of isosinglet down-type quarks $b'_L,b'_R$, with the mass of few TeV,  
which has rather strongly mixed with the first family $V_{ub'}\simeq 0.04$.  
However, apart of the persistent question "who has ordered that?", it has some 
rather unnatural features. In particular, in order to avoid strong flavor changing effects 
in Kaon physics etc.,  the 4th quark $b'$ should have weaker mixings with 2nd and 3rd families 
than with the first one. Perhaps such a situation is possible by some conspiracies, 
however a priori it looks rather weird. 

Alternatively, an additional effective operator contributing to muon decay in positive interference 
with the Standard Model contribution can restore unitarity.
In this case the Fermi constant wolud be slightly different from muon decay constant, 
$G_F = G_\mu/(1+\delta_\mu)$, where $\delta_\mu \simeq 7\times 10^{-4}$ 
would suffice for restoring the unitarity.   
Namely, the values of $V_{us}$ and $V_{ud}$ (which are normally extracted by assuming $G_F=G\mu$)
are shifted by a factor $1+\delta_\mu$ while their ratio is not affected.
The needed effective operator can be mediated by a 
flavor changing boson related to a gauge horizontal 
symmetry $SU(3)_\ell$ acting between the three lepton families, 
which symmetry is spontaneously broken at the scale of $6-7$ TeV. 

Considering the gauge symmetry group $SU(3)_\ell\times SU(3)_e$ 
acting on left-handed and right-handed leptons respectively, one can get 
a natural understanding on the origin of the mass hierarchy among 
charged leptons and large mixing angles for neutrinos 
which is related to the pattern of spontaneous breaking of the symmetry. 
Interestingly, despite the fact that these gauge bosons are have flavor--changing couplings 
with the leptons, their exchanges do not induce dramatic LFV effects as decays 
$\mu\to 3e$, $\tau \to 3\mu$ etc., which can be kept under control 
thanks to approximate custodial symmetry.
 
 Analogously, one can consider gauge symmetry  
 $SU(3)_Q \times SU(3)_u \times SU(3)_d$ between the quark families.   
  Its breaking pattern can be at the origin of the quark mass and mixing hierarchy, 
and the flavor-changing  gauge bosons of  $SU(3)_Q$ can contribute to hadronic decays of 
kaons, hyperons, etc. 
 In supersymmetric extension of the SM, the chiral gauge symmetries 
 $SU(3)_\ell\times SU(3)_e$ for leptons and $SU(3)_Q \times SU(3)_u \times SU(3)_d$ 
for quarks can be also motivated as a natural  possibility of the realizing 
the minimal flavor violation scenario \cite{PLB98,MFV}.  

One interesting possibility, discussed in Ref. \cite{Benedetta}, is that 
these flavor gauge symmetries are common symmetries between the ordinary 
and mirror particle scoters, which is also motivated to the possibility of cancellation 
of triangle anomalies of gauge $SU(3)$ factors between the ordinary and mirror 
particles \cite{PLB98}. Mirror matter is also a viable candidate for dark matter 
(see e.g. reviews \cite{Alice}). 
Since flavor gauge bosons are messengers between the two sectors,
 then they are a portal for direct detection of mirror dark matter \cite{Cerulli} 
 but also they mediate new flavor violating 
 phenomena such as muonium--mirror muonium, kaon--mirror kaon 
 oscillations \cite{Benedetta}.  

\bigskip 

\noindent 
{\bf Acknowledgements.}
The work of Z.B. was supported by the research grant 
"The Dark Universe: A Synergic Multimessenger Approach" No. 2017X7X85K 
under the program PRIN 2017 funded by the Ministero dell'Istruzione, Universit\`a e della Ricerca (MIUR). 
The work of R.B. and Z.B. was supported in part by Shota Rustaveli National Science Foundation 
(SRNSF) of Georgia, grant DI-18-335/New Theoretical Models for Dark Matter Exploration.

 \end{document}